\documentclass{pas}
\usepackage{aas_macros}

\usepackage{multirow}
\usepackage{natbib}
\usepackage{amsmath}
\usepackage{bm}
\newcommand\orcids[1]{\href{https://orcid.org/#1}{\includegraphics[height=8.5pt,trim={-6pt 0 -6pt 0},clip]{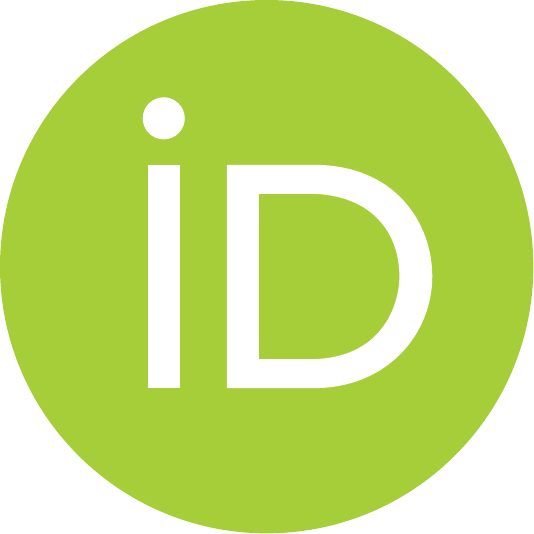}}}

\begin{document}

\lefttitle{Publications of the Astronomical Society of Australia}
\righttitle{Stace et al.}

\jnlPage{1}{4}
\jnlDoiYr{2025}
\doival{10.1017/pasa.xxxx.xx}

\articletitt{Research Paper}

\title{Multi-Resonant-Line Radiative Transfer: \\ Lyman-Alpha Fine Structure and Deuterium Coupling}

\author{Ethan Stace\orcids{0009-0000-5101-5657}$^{1}$, Aaron Smith\orcids{0000-0002-2838-9033}$^{2}$, Kevin Lorinc\orcids{0009-0005-3827-8774}$^{2}$, and Olof Nebrin\orcids{0000-0003-3877-360X}$^{3}$}

\affil{$^{1}$Department of Physics, University of Florida, Gainesville, Florida 32611, USA}

\affil{$^{2}$Department of Physics, The University of Texas at Dallas, Richardson, Texas 75080, USA}

\affil{$^{3}$Department of Astronomy \& Oskar Klein Centre for Cosmoparticle Physics, AlbaNova, Stockholm University, SE-106 91 Stockholm, Sweden }

\corresp{Smith. A. Email: \href{mailto:asmith@utdallas.edu}{asmith@utdallas.edu}}



\begin{abstract}
    Resonance lines encode rich information about astrophysical sources and their environments, yet fully analytic treatments of multi-line radiative transfer remain almost entirely unexplored. We present exact, closed-form solutions for steady-state resonant-line radiative transfer in ``V-shaped'' atomic networks, where a single ground state couples to multiple transitions. Starting from the full angle-dependent transfer equation, we generalise absorption and emission coefficients to an arbitrary number of lines, derive a modified Fokker--Planck expansion of the frequency-redistribution integral, and use a judicious change of variables to reduce the problem to a Helmholtz equation with point-like sources in frequency space. This transformation admits analytic solutions for arbitrary sets of lines with fixed frequency offsets and strengths in both slab and spherical geometries. We implement V-shaped line networks in the \textsc{colt} Monte Carlo radiative transfer code and find excellent agreement with the analytic predictions across a wide range of line separations, optical depths, and damping parameters, establishing our solutions as stringent validation benchmarks. For concrete applications related to the Lyman-alpha (Ly$\alpha$) transition of neutral hydrogen, we examine how fine-structure splitting and deuterium injection modify the emergent spectra, internal radiation field, and radiative force multiplier. We show that these effects leave previous conclusions about Ly$\alpha$ feedback in the early universe essentially unchanged. Even when direct observational diagnostics are subtle, our framework provides novel analytic and numerical insights into coupled resonance-line transport and facilitates progress in general modelling of multi-line radiative transfer in diverse astrophysical settings.
\end{abstract}


\maketitle

\section{Introduction}\label{sec:intro}

Radiative transfer (RT) plays a central role in astrophysics by describing how light propagates and interacts with astrophysical media. The RT equation (RTE) provides a powerful framework for interpreting observations and modelling multi-scale radiation-induced phenomena, but solving it exactly is often intractable due to the complexity of scattering processes, especially for general three-dimensional geometries \citep{Haworth2018}. As a result, various approximations are often made to obtain analytical solutions. Examples include assuming a steady state, integrating over all photon frequencies, or simplifying the scattering physics involved \citep{Allakhverdian2023}. These approximations trade some physical fidelity for mathematical simplifications, and careful choices must be made to capture the dominant effects.

One application with particularly rich physics is the RT of the hydrogen Lyman-alpha (Ly$\alpha$) line. Ly$\alpha$ is intrinsically bright due to the pervasive nature of photo-ionisation producing recombination events, each of which has probability $P(\text{Ly}\alpha) \approx 0.68$ of yielding a Ly$\alpha$ photon with a high cross-section for scattering with neutral hydrogen \citep[e.g.][]{Dijkstra2019}. Understanding any mechanisms impacting Ly$\alpha$ RT in the local and high-redshift universe is therefore critical for interpreting observations of star-forming galaxies and the intervening intergalactic medium (IGM). However, Ly$\alpha$ photons undergo multiple resonant scatterings, making their propagation highly diffusive in both physical and frequency space. This process shapes the emergent Ly$\alpha$ spectra and surface brightness profiles, complicating any direct connection between observed signals and their source properties. Analytical models, even if idealised, are invaluable for clarifying these effects and guiding interpretations of Ly$\alpha$ observations and simulations.

Over the years, an increasing number of analytic solutions for Ly$\alpha$ have been extensively developed under various simplifying assumptions. An important milestone was the adoption of the Fokker--Planck approximation for partial frequency redistribution, allowing closed-form solutions for infinite plane-parallel slab geometry with central point and uniform sources by \citet{Harrington1973} and \citet{Neufeld1990}, building on work done by \citet{Unno1955}. Then, configurations with spherical and cubic symmetry were derived by \citet{Dijkstra2006} and \citet{Tasitsiomi2006}, respectively, extending the theory to different geometries. These classical solutions assumed static, homogeneous media and yielded characteristic double-peaked Ly$\alpha$ line profiles in the emergent spectrum. While the basic behaviour is well understood, significant progress is still being made with explicit formulas for the internal radiation field within the medium by \citet{LaoSmith2020} and \citet{SeonKim2020}, whereas \citet{McClellan2022} incorporated time-dependent, impulsive emission sources into the analytic framework. Variations in density and source distributions were explored by \citet{LaoSmith2020} for power-law density and emissivity gradients, while velocity gradients have been included by \citet{Nebrin2025} and \citet{Smith2025} to account for bulk flows, e.g. outflows. Most comprehensively, \citet{Nebrin2025} presented a generalised series solution that combines several additional physical effects, such as macroscopic velocity fields, atomic recoil, Ly$\alpha$ photon destruction via $2\text{p} \to 2\text{s}$ transitions, small-scale density fluctuations induced by turbulence, and continuum absorption/scattering, all within a single model. Each of these advancements has improved the realism of analytic RT models. However, despite this extensive progress, all of the above studies are limited to a single spectral line at a time. In other words, existing analytic solutions describe strong line RT in isolation, without accounting for the possibility of multiple resonant transitions interacting or overlapping in a common medium.

Modelling the transfer of multiple spectral lines simultaneously is a natural next step with broad implications. In the context of Ly$\alpha$ RT, considering multiple lines can capture processes such as fine-structure splitting and the Wouthuysen--Field effect (the coupling of Ly$\alpha$ photons with the 21-cm hyperfine transition). More generally, multi-line RT is essential in environments where different lines overlap or feed into each other. Notable examples include the rich Lyman-series cascades in recombining gas, the Balmer lines in dense nebulae, and line-driven outflows around active galactic nuclei (AGNs), accretion disk winds, and winds around massive stars and supernovae. In such systems, radiation in one transition can alter the population of atoms or ions in excited states, thereby affecting the optical depth and emission in another transition. Numerical simulations have been used to model multi-line RT \citep{Lucy2002, SIROCCO, DruettZharkova2018, Chang2025}. To make this feasible, Monte Carlo codes typically employ the ``macro-atom'' formalism or Sobolev approximation to handle the increased complexity, but often by sacrificing detailed treatment of the frequency redistribution or assuming instant re-emission in a different line. In some contexts, iterative schemes have been used to solve coupled RTEs for all optically thick lines of hydrogen simultaneously. These approaches have provided valuable insights, but often discard the spectral memory of previous scatterings. To date, we are not aware of any closed-form analytic solution that treats more than one resonant line at a time with detailed partial frequency redistribution physics.

In this paper, we present the first analytic solution of the RTE that self-consistently accounts for multiple resonant lines simultaneously. We focus on the simplest non-trivial case of a ``V-shaped'' two-level atom network, in which a single ground state is coupled to multiple independent excited states and their corresponding atomic line transitions. This scenario represents the least-complex case of a broader class of multi-line networks, where one would consider higher-order transitions and radiative cascades, greatly increasing the scattering physics complexity. Still, the V-shaped system already captures phenomena like Ly$\alpha$ fine-structure doublet splitting and hydrogen--deuterium line blending. Our analytic solution is obtained under the same assumptions as used in single-line Ly$\alpha$ solutions. We also apply the diffusion approximation for frequency redistribution, which treats each scattering as an effectively instantaneous, small frequency change event. Within this coupled framework, we derive a closed-form expression for the stead-state spectral intensity as a function of position and frequency, for both the internal radiation field and the emergent flux escaping the medium. An important development enabling this solution is the generalisation of the Fokker--Planck formulation of the redistribution function for multiple lines, a powerful result that will be derived more generally in a future paper.

To validate our analytic results and illustrate their utility, we have also developed a complementary numerical approach. We modified the Monte Carlo based Cosmic Ly$\alpha$ Transfer code \citep[\textsc{colt};][]{Smith2015} to handle V-shaped multi-line networks with accurate frequency redistribution. In essence, our code allows photons to scatter into any of the available upper-state transitions, with the choice of the line determined probabilistically by the local relative line optical depths. Using this tool for explicit multi-line resonant scattering physics, we perform simulations of Ly$\alpha$ fine-structure transfer and deuterium--hydrogen mixed media, and we show that the simulation results agree quantitatively with the predictions of our analytic solution. This cross-validation gives us confidence in the correctness of the theory, highlighting the improved physical accuracy as both analytic and numerical methods involve non-trivial generalisations.

The rest of this paper is organised as follows. In Section~\ref{sec:equations}, we introduce the governing RTE and generalisations of the many variables in the RTE, such as the definition of a multi-line dimensionless frequency coordinate and a combined absorption profile. In Section~\ref{sec:analytics}, we present the full analytical solution for both internal and emergent spectra in the case of the V-shaped network, with the details of the derivation and other details provided in the appendices. In Section~\ref{sec:simulations}, we outline the modifications made to \textsc{colt} to simulate V-shaped scattering, and we demonstrate its accuracy with the analytical solutions developed in Section~\ref{sec:analytics} for various test cases using arbitrary atomic parameters. In Section~\ref{sec:applications}, we apply our new tools to two relevant astrophysical scenarios of Ly$\alpha$ fine-structure and deuterium injection. These examples show how multi-line effects manifest in realistic conditions and confirm that our models reproduce known results from the literature in the appropriate limits. Finally, in Section~\ref{sec:discussion}, we summarise our findings, discuss the astrophysical implications of multi-line RT, and outline potential directions for future work, including the incorporation of additional physics and the extension to more complex multi-level atomic networks. Appendix~\ref{ap:analytical solution} contains the derivation of the analytical solutions of this text, and Appendix~\ref{ap:x tilde change of variables} contains a discussion on the required change of frequency variables. Finally, Appendix~\ref{ap:parameter_table} contains a table of the parameters used for each figure for reproducibility purposes and to explain line asymmetries.

\section{Governing Equations}\label{sec:equations}

The general form of the steady-state RTE in static media is:
\begin{equation} \label{eq:RTE nu}
  \bm{n} \bm{\cdot} \bm{\nabla} I_\nu = j_\nu - k_\nu I_\nu + \iint I_{\nu'} (kR)_{\nu', \bm{n}' \rightarrow \nu, \bm{n}}\,\text{d}\Omega' \text{d}\nu' \, ,
\end{equation}
where $I_\nu(\bm{r}, \bm{n})$ is the specific intensity (for frequency $\nu$ at position $\bm{r}$ travelling in direction $\bm{n}$), $j_\nu(\bm{r})$ and $k_\nu(\bm{r})$ denote the emission and absorption coefficients, and the last term accounts for frequency redistribution due to partially coherent scattering \citep{Dijkstra2019}. The term $(k R)$ relates to the redistribution function $R_{\nu', \bm{n}' \rightarrow \nu, \bm{n}}$, which is the differential probability per unit initial photon frequency $\nu'$ and per unit initial directional solid angle $\Omega'$ that the scattering of such a photon travelling in direction $\bm{n}'$ would place the scattered photon at frequency $\nu$ and directional unit vector $\bm{n}$ \citep{Hummer1962}. The notational convention of $(k R)$ was chosen for convenience with multi-line extensions and has the same units as $k_\nu$. Physically, $R$ encodes the quantum mechanical scattering processes (e.g. Doppler shifts, natural broadening, recoil, etc.), and couples the RTE across different frequencies.

The frequency dependence of the absorption coefficient is given by the Voigt profile $\phi_\text{Voigt}$ centred on the line resonance. For convenience we define the Hjerting--Voigt function $H(x, a) = \sqrt{\pi} \Delta \nu_\text{D} \phi_\text{Voigt}(\nu)$ as the dimensionless convolution of Lorentzian and Maxwellian distributions,
\begin{equation} \label{eq:H}
  H(x, a) = \frac{a}{\pi} \int_{-\infty}^\infty \frac{e^{-y^2}\text{d}y}{a^2+(y-x)^2} \approx
    \begin{cases}
      e^{-x^2} & \quad \text{`core'} \\
      {\displaystyle \frac{a}{\sqrt{\pi} x^2} } & \quad \text{`wing'}
    \end{cases}
\end{equation}
where the `damping parameter' is $a \equiv \Delta \nu_\text{L} /2 \Delta \nu_\text{D}$,  $\Delta \nu_\text{L}$ is the natural line width of the line, $\Delta \nu_\text{D} \equiv \nu_0\,v_\text{th} / c$ is the Doppler width, and $v_{\text{th}} = (2\,k_{\text{B}}T/m)^{1/2}$ is the thermal velocity at temperature $T$ for carrier mass $m$, such that $x = (\nu-\nu_0)/\Delta\nu_\text{D}$ measures the frequency offset from line centre in Doppler units.

For a multi-line system, we generalise the notation to arrive at a compact modified RTE. With $N$ lines, we have multiple distinct resonance frequencies $\nu_{0, i}$ for $i \in \{1,\, ...,\, N\}$. We denote the dimensionless frequency variables for each individual line as $x_i$ and redefine $x$ as the unified dimensionless frequency variable as
\begin{equation} \label{eq:x}
    x_i \equiv \frac{\nu - \nu_{0,i}}{\Delta\nu_{\text{D},i}} = \mu_i(x-\Delta_i) \quad \text{and} \quad x \equiv \frac{\nu - \nu_0}{\Delta\nu_\text{D}} = \frac{1}{N} \sum_{i = 1}^{N}x_i \, ,
\end{equation}
respectively, where we defined $\mu_i = \Delta\nu_\text{D} / \Delta\nu_{\text{D},i}$. Here $\nu_0$ and $\Delta\nu_\text{D}$ can be arbitrary reference values but for concreteness we may adopt the mean line frequency $\nu_0 \equiv \sum_{i=1}^N\nu_{0,i}/N$ and Doppler width $\Delta\nu_\text{D} = \sum_{i = 1}^N \Delta\nu_{\text{D}, i} / N$. The dimensionless offset of each line in this common $x$ coordinate is $\Delta_i \equiv (\nu_{0, i} - \nu_0) / \Delta\nu_\text{D}$. In practice, $\mu_i$ departs from unity when lines have relatively large separations or different carrier masses as $\Delta\nu_{\text{D}, i} \propto \nu_{0,i} / \sqrt{m_i}$. Throughout this work we will use $x$ as the master frequency variable and treat each line's profile in shifted coordinates $x_i$.

The total absorption coefficient is the sum of contributions from all lines \citep{AhnLee2015, SeonKim2020}:
\begin{align} \label{eq:k_x}
  k_x &\equiv \sum_{i=1}^N k_{x,i} = \sum_{i=1}^N n_i\,\sigma_{0, i}\,H(a_i, x_i) \nonumber \\
  &= \sum_{i = 1}^N n_i\,f_{lu,i}\,\frac{\pi\,e^2}{m_e\,c\,\Delta\nu_{\text{D},i}}\,H(a_i, x_i) \equiv k(\bm{r})\bar{H}(x) \, ,
\end{align}
where $n_i$ is the number density of atoms in the lower level of a given transition and $\sigma_{0, i}$ is the line centre cross-section defined in terms of the oscillator strength $f_{lu,i}$. In the last equality, we have factored out the position dependence into $k(\bm{r}) = \sum_{i = 1}^Nn_i\sigma_{0, i}$, the total line-centre absorption coefficient of all lines, and defined a multi-lined Voigt profile that encodes the relative strength of each line within dimensionless weights to represent each contribution:
\begin{equation} \label{eq:H}
  \bar{H}(x) \equiv \sum_{i=1}^N \omega_i H(a_i,x_i)
  \quad \text{where} \quad \omega_i \equiv \frac{n_i\,f_{lu, i}} {\Delta\nu_{\text{D}, i}} \Big/ \sum_{i=1}^N  \frac{n_i\,f_{lu, i}} {\Delta\nu_{\text{D}, i}} \, .
\end{equation}
By construction, the weights are normalised to unity, $\sum_{i=1}^N \omega_i = 1$, and reduce to the simple abundance ratio of level populations in the case of lines with similar oscillator strengths and thermal widths. We later simplify each of these coefficients further for the specific applications explored in this paper.

We must extend the scattering term in Eq.~(\ref{eq:RTE nu}) to account for multiple lines. For V-shaped networks, the redistribution function can be written as a sum of contributions from each line scattering back into the same line. Adopting the common $x$ coordinate for the integrand but with line-specific parameters gives:
\begin{align}
    &\iint I_{x'} (kR)_{x', \bm{n}' \rightarrow x, \bm{n}}\,\text{d}x' \text{d}\Omega' \longrightarrow \notag \\
    & \, k(\bm{r}) \sum_{i = 1}^N \omega_i \iint I_{x'} H(a_i, x'_i) R_{x' - \Delta_i, \bm{n}' \rightarrow x - \Delta_i, \bm{n}}\,\text{d}x'\text{d}\Omega' \, .
\end{align}
We prefer this form over an equivalent sum of shifted frequency integrals to unify Fokker--Planck terms.

\begin{figure}
    \centering
    \includegraphics[width=0.475\textwidth]{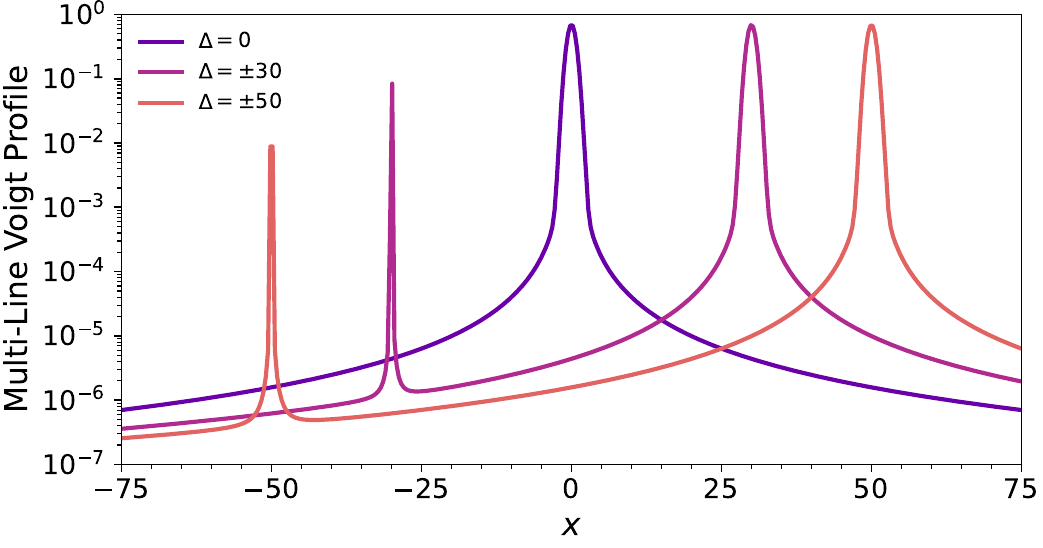}
    \caption{Multi-line Voigt profile $\bar{H}(x)$ from Eq.~(\ref{eq:H}) for a two-line system with asymmetric line strengths. The stronger line dominates except in the immediate vicinity of the weaker line. The normalisation also affects the weak-line heights.}
    \label{fig:H_bar}
\end{figure}

\section{Analytical Solutions} \label{sec:analytics}

Using the formulations above, we now derive analytical solutions to the RTE for multi-line V-shaped networks. The full derivation can be found in Appendix~\ref{ap:analytical solution}; here we summarise the results and provide the final expressions for the specific intensity. We focus primarily on the simplest case of two resonant ground state transitions ($N=2$), generalising afterwards. For clarity, we follow the notation of \citet{LaoSmith2020}, and write the general multi-line angle-averaged intensity $J$ as governed by
\begin{equation} 
  \frac{1}{k(\bm{r})} \bm{\nabla}\!\bm{\cdot}\!\left(\frac{\bm{\nabla} J}{k(\bm{r})} \right)\!+\!\frac{3}{2} \bar{H}(x) \frac{\partial}{\partial x}\!\left(\!\hat{H}(x) \frac{\partial J}{\partial x} \right) = -\frac{3\mathcal{L}}{4\pi}\!\frac{\eta(\bm{r})}{k(\bm{r})}\!\frac{\bar{H}\tilde{H}}{\sqrt{\pi}} \, ,
\end{equation} 
where we defined further combinations of individual Voigt profiles
\begin{equation}
    \tilde{H}(x) = \sum_{i=1}^{N}\omega_i\mu_iH(x_i) \quad 
    \text{and} \quad     
    \hat{H}(x) = \sum_{i=1}^{N}\frac{\omega_i}{\mu_i^2}H(x_i) \, ,
\end{equation}
which are used to make the equations analytically tractable, and are justified by the following frequency change of variables:
\begin{equation} \label{eq:change_of_variables_true}
    \text{d}\tilde{x} = \sqrt{\frac{2}{3}} \frac{\text{d}x}{\tau_0(\bar{H}\hat{H})^{1/2}} \quad \implies \quad \tilde{x} = \int_0^{x}\sqrt{\frac{2}{3}}\frac{\text{d}x'}{\tau_0(\bar{H}\hat{H})^{1/2}} \, .
\end{equation}
This yields a solution for any general $N$-line V-shaped network:
\begin{align} \label{eq:superposition1}
    &J(\tilde{\bm{r}}, \tilde{x}) = \sum_{i = 1}^N \omega_i\mu_i J_i(\tilde{\bm{r}}, \tilde{x}-\tilde{\Delta}_i) \, ,\quad \text{where} \quad \\
    &J_i(\tilde{\bm{r}},\tilde{x}-\tilde{\Delta}_i) = \frac{\mathcal{L} \sqrt{6}}{8 \pi} \sum_{n=1}^\infty \frac{Q_{n}}{\lambda_{n}} \vartheta_n(\tilde{\bm{r}}) e^{-\lambda_{n}|\tilde{x}-\tilde{\Delta}_i|} \, ,
\end{align}
where $\mathcal{L}$ is the total luminosity of the source embedded in the medium, $\vartheta_n(\bm{r})$ and $\lambda_n$ are the spatial eigenfunctions and eigenvalues obtained from the diffusion equation depending on the geometry and boundary conditions, and $Q_n$ is a geometrical coupling coefficient defined in Appendix~\ref{ap:analytical solution}. The tildes on $\tilde{\bm{r}}$ and $\tilde{x}$ indicate that these coordinates have been transformed into dimensionless rescaled coordinates, i.e. the position normalised to the optical depth and a line-integrated frequency defined and calculated in Appendix~\ref{ap:x tilde change of variables} that greatly simplifies the differential equation. When compared with the single line solutions of \citet{LaoSmith2020}, the multi-line solution is simply a linear combination of single-line solutions centred on each line, i.e. a superposition weighted by the relative line strength $\omega_i$ and scaled by $\mu_i$ to account for differences in thermal width. This is a gratifyingly simple result, as the complicated coupling between lines can be encapsulated by known results, provided that it is possible to work out the common transformed frequency variable $\tilde{x}$.

\begin{figure}
    \centering
    \includegraphics[width=\columnwidth]{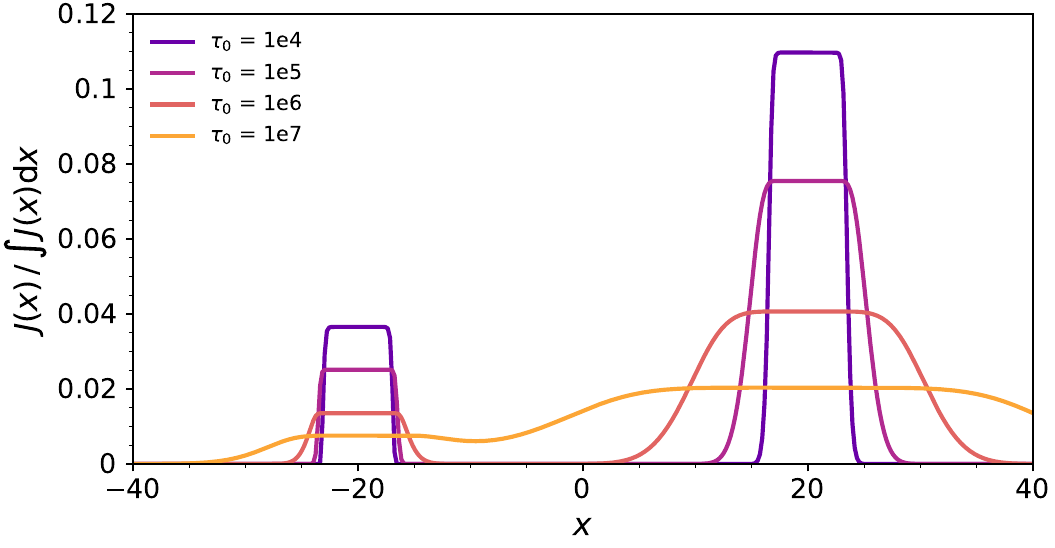}
    \includegraphics[width=\columnwidth]{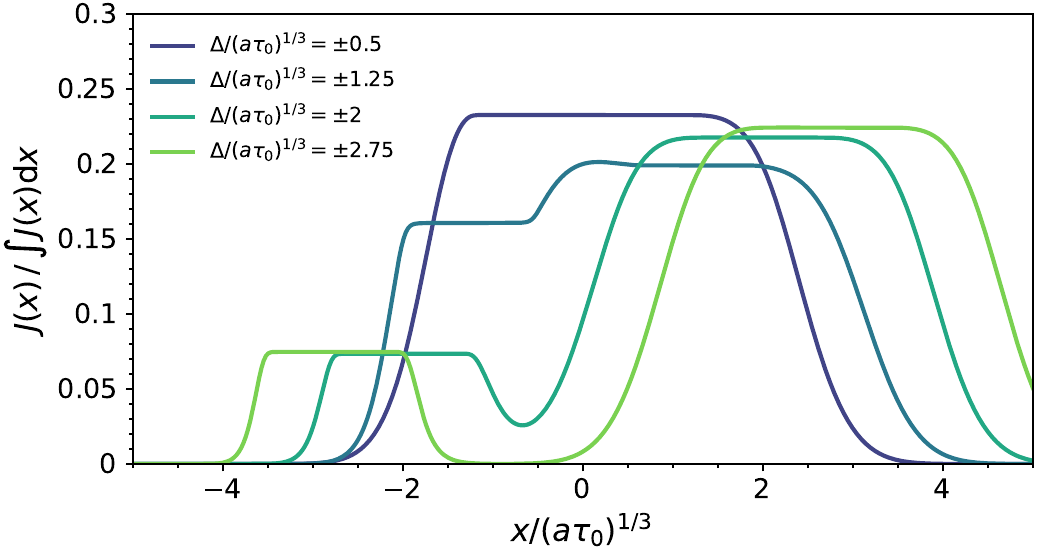}
    \caption{A two-line V-shaped network with various optical depths and line separations. The asymmetry in chosen line parameters ($\omega_1 = 1/3$ and $\omega_2 = 2/3$) causes a difference between the peaks, with the second line being much stronger than the first. \textit{Upper Panel:} Internal spectrum for a plane-parallel slab at $\tilde{r} = 0.5$, changing the optical depth over $\tau_0 \in \{10^4, 10^5, 10^6, 10^7\}$ at a temperature of $T = 10^4$\,K, based on Eq.~(\ref{eq:PS-interior-solution-slab}). Higher optical depth enables more diffusion across frequency space, since photons must typically reach a critical frequency of order $x_\text{esc} \sim (a \tau_0)^{1/3}$ before escape, increasing the likelihood of coupling between nearby lines. \textit{Lower Panel:} Internal spectrum for a spherical cloud geometry with different line separations from the centre as parametrised by $\Delta \in \pm \{0.5, 1.25, 2, 2.75\}\,(a\tau_0)^{1/3}$ at $\tilde{r} = 0.5$, based on Eq.~(\ref{eq:PS-interior-solution-sphere}). As the spacing increases the solution separates into two distinct profiles, decoupling from each other. While there is still significant overlap the relative heights are similar due to mixing between lines.}
    \label{fig:internal}
\end{figure}

To make these results concrete, we now specialise to commonly assumed geometries and source configurations. We consider (\textit{i}) an infinite homogeneous slab of height $Z$ with a symmetric source at the mid-plane, and (\textit{ii}) a uniform sphere of radius $R$ with a central source, both characterised by centre-to-edge optical depths of $\tau_0$. These are the same setups originally considered by \citet{Harrington1973}, \citet{Neufeld1990}, and \citet{Dijkstra2006} for single-line transfer, and therefore provide benchmark cases to understand how the presence of a second line alters classic solutions.

For slab geometry, we adopt the dimensionless spatial coordinate $\tilde{z} = z / Z$ and place a delta-function source at $z=0$. The solution for the multi-line internal angle-averaged intensity is
\begin{equation} \label{eq:PS-interior-solution-slab}
    J(\tilde{z},\tilde{x}) = \frac{\mathcal{L}\sqrt{6}}{8\pi^2} \sum_{i = 1}^N \omega_i\mu_i\,\text{tanh}^{-1}\!\left\{\text{cos}\!\left(\frac{\pi\tilde{z}}{2}\right)\text{sech}\!\left[\frac{\pi(\tilde{x}-\tilde{\Delta}_i)}{2}\right]\right\} \, ,
\end{equation}
and the emergent (surface) spectrum from the slab is obtained by evaluating the boundary condition at $\tilde{z} = 1$, which yields
\begin{equation} \label{eq:PS-emergent-solution-slab}
    J(\tilde{x}) = \frac{\mathcal{L}\sqrt{6}}{16\pi f\tau_0\bar{H}(\tilde{x})} \sum_{i = 1}^N \omega_i\mu_i\,\text{sech}\!\left[\frac{\pi(\tilde{x}-\tilde{\Delta}_i)}{2}\right] \, ,
\end{equation}
where $f$ is an order unity constant determined by the choice of boundary conditions. These results reduce to the standard single-line solutions when one line dominates, i.e., $\omega_1 \approx 1$ and $\tilde{\Delta}_1 \approx 0$. The general behaviour of the $N=2$ slab solution is illustrated in the top panel of Fig.~\ref{fig:internal}, showing that as the optical depth increases, the intensity profile broadens and the two lines begin to blend together into a single, wider profile once the line separation is within the reach of the resonant broadening, i.e. $\Delta \lesssim (a \tau_0)^{1/3}$. For high enough optical depth, photons collectively scatter far enough in frequency to bridge the gap between the two resonances, making the RT resemble that of a single effective line with combined optical depth.

For spherical geometry, we adopt $\tilde{r} = r / R$ and place a delta-function source at $r=0$. The multi-line internal solution is
\begin{equation} \label{eq:PS-interior-solution-sphere}
  J(\tilde{r},\tilde{x}) = \frac{\mathcal{L} \sqrt{6}}{32 \pi^2 R^2 \tilde{r}} \sum_{i = 1}^N \frac{\omega_i\mu_i\,\sin(\pi \tilde{r})}{\cosh[\pi(\tilde{x}-\tilde{\Delta}_i)] - \cos(\pi \tilde{r})} \, ,
\end{equation}
and the emergent intensity is
\begin{equation} \label{eq:PS-emergent-solution-sphere}
  J(\tilde{x}) = \frac{\mathcal{L}\sqrt{6}}{64 \pi R^2 f \tau_0 \bar{H}(\tilde{x})} \sum_{i = 1}^N \omega_i\mu_i\,\text{sech}^2\!\left[\frac{\pi(\tilde{x}-\tilde{\Delta}_i)}{2}\right] \, .
\end{equation}

In the bottom panel of Fig.~\ref{fig:internal}, we show the internal spectrum at an arbitrary radius $\tilde{r} = 0.5$ for a two-line system. Again, when the separation between the lines is small compared to the diffusion width of each line, the two peaks overlap and blend into each other. Conversely, when the lines are well separated (relative to their widths), even the internal spectrum shows two distinct diffusion peaks, quantifiable by the separability condition $|\Delta_2 - \Delta_1| \gg (a \tau_0)^{1/3}$. More precisely, the full-width at half-maximum (FWHM) of a single-line profile can be found using equation~(90) in \citet{LaoSmith2020} in $\tilde{x}$ space, converting to $x$ space using Eq.~(\ref{eq:change-of-variables}) and the approximation $\bar{H}(x) \approx \omega_i H_i(x_i, a_i)$ valid around the neighbourhood $x \approx \Delta_i$, which gives $\text{FWHM}_i \approx 1.67\,(\omega_i a_i \tau_0)^{1/3}$. Overlap occurs when $|\Delta_i - \Delta_j| \lesssim \text{FWHM}_i + \text{FWHM}_j$. For example, in Fig.~\ref{fig:internal}, we have $(\text{FWHM}_1 + \text{FWHM}_2) / (a\tau_0)^{1/3} = 2.57$, indicating a significant overlap for $\Delta \leq 1.3\,(a\tau_0)^{1/3}$. A more subtle variable that also impacts line separation is temperature, as we can see from $\Delta_i \propto T^{-1/2}$ and $(a_i\tau_0)^{1/3} \propto T^{-1/3}$, and so at higher temperatures the distributions merge together more easily.

\begin{figure}
    \centering
    \includegraphics[width=0.475\textwidth]{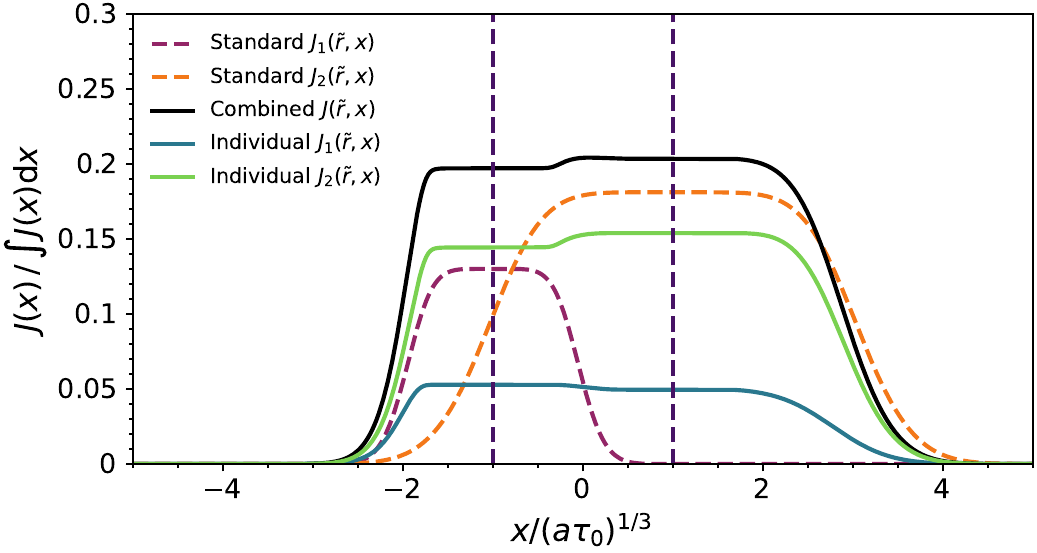}
    \caption{Internal spectrum $(\tilde{r} = 0.5)$ contributions of individual lines, i.e. $J_1$ and $J_2$ from Eq.~(\ref{eq:PS-interior-solution-sphere}),
    in comparison to standard single-line solutions. We emphasise that the terms exhibit qualitatively different behaviour as only the sum $J$ is physically meaningful across the full spectrum, rather than $J_1$ and $J_2$ in isolation.}
    \label{fig:J_2_individual}
\end{figure}

We stress that while Eq.~(\ref{eq:superposition1}) suggests a simple superposition of single-line solutions, one must be careful in interpreting the individual terms. The transformation $\tilde{x} = \tilde{x}(x)$ depends on the combined profile $\bar{H}(x)$ (see Appendix~\ref{ap:x tilde change of variables}), which means that in regions of frequency space dominated by one line, the coordinate mapping for the other line can become highly non-linear. An illustration of this phenomenon is given in Fig.~\ref{fig:J_2_individual}, where we present the contributions of each of the lines from Eq.~(\ref{eq:PS-interior-solution-sphere}). Individual terms do not match the standard single-line solutions derived in \citet{LaoSmith2020} centred on the same frequency. In fact, the $J_2$ term deviates strongly because $\text{d}\tilde{x}/\text{d}x \rightarrow 0$ around $x \approx \Delta_1$. Therefore, these individual terms are not to be used in isolation, only the sum $J$ is physically meaningful across all frequencies. To better understand the behaviour of $\tilde{x}$, see Appendix~\ref{ap:x tilde change of variables} for more details. Finally, we note that many other quantities typically derived also have similar superposition relations, i.e. energy density.

\section{Numerical Simulations}
\label{sec:simulations}

To check the validity of our analytic solutions, we implemented a new flexible multi-line resonant scattering Monte Carlo radiative transfer (MCRT) module within the Cosmic Ly$\alpha$ Transfer code \citep[\textsc{colt};][for public code access and documentation see \href{https://colt.readthedocs.io}{\texttt{colt.readthedocs.io}}]{Smith2015, Smith2019, Smith2022, McClymont2025COLT}. The \textsc{colt} code is well-tested for Ly$\alpha$ and we extend this functionality to handle $N$-line V-shaped atomic networks as in Section~\ref{sec:equations}. 
The modifications are similar to those described by \citet{SeonKim2020} for including the Wouthuysen--Field effect, i.e. including fine-structure for Ly$\alpha$--21\,cm coupling in MCRT frameworks. In essence, the algorithm proceeds as follows: to trigger scattering events we draw a random optical depth $\tau_\text{scat} = -\ln\xi$, where $\xi$ is a random univariate, from the exponential distribution $e^{-\tau}$ and calculate the scattering distance by equating this to the total traversed optical depth from a minimal set of nearby lines. We then draw another random number to determine which line the photon actually interacts with. This is done by comparing the relative contributions of each line to the total optical depth, as the probability to scatter into line $i$ is $P_i = \tau_i / \sum \tau_i$, where the $\tau_i$ are partial optical depths for each line. After selecting the line, we treat the event as a standard resonant scattering in that transition, drawing the outgoing direction and frequency based on the atomic properties of the line \citep[either artificial or real data, e.g.][]{Sansonetti} and the corresponding phase and redistribution functions. This sampling captures the true microphysics, where photons interact with one atom at a time. In this paper, we focus on V-shaped networks, assuming all atoms are in the ground state, which is valid where the radiation field is not intense enough to significantly populate excited states. Future work will explore non-LTE resonant pumping effects by iterating between transport and level population statistics. We emphasise that our MCRT implementation fully retains frequency-dependent scattering physics without Sobolev-like or escape probability approximations; thus, it serves as a robust numerical check on our analytic diffusion solutions. We refer to a specific set of levels, allowed transitions, and their corresponding atomic data as a ``network''. We feature three networks in this paper: (\textit{i}) an arbitrary two-line V-shaped network with variable line separation, (\textit{ii}) the hydrogen Ly$\alpha$ fine-structure doublet, and (\textit{iii}) a Deuterium injected Ly$\alpha$ network.

For simplicity, in all simulations we adopt the same phase functions as most Ly$\alpha$ codes \citep{Dijkstra2006}, i.e. dipole scattering in the wings with modifications in the core. While emergent spectra are in fact nearly independent of the phase function because of multiple scatterings, in principle, one could extend the networks to have dynamic phase functions that vary with incoming frequency for a self-consistent quantum mechanical treatment or polarisation using a Stokes vector approach as in \citet{SeonSong2022}, but this is beyond the scope of the current study.

For the applications of this paper we use a static, uniform-density configuration with either slab or spherical geometry. We inject photons at the centre of the medium (point source) with an initial frequency chosen such that the photon is equally likely to be in any of the lines, i.e. we match line $i$ with probability weight $\omega_i$. For convenience, we calculate $a\tau_0$ using the $a_i$ from the line with lower frequency, unless otherwise specified. In many cases, the $a_i$ used can be arbitrary, as $a\tau_0$ is simply a frequency diffusion scaling factor. For concreteness we list the specific parameter values used in each figure in Appendix~\ref{ap:parameter_table} for reproducibility.

For optically-thick tests, we have confirmed that the standard constant core-skipping algorithm works well as long as $x_\text{crit}$ for each independent line is not too large, where $x_\text{crit}$ is the minimum perpendicular velocity drawn from a volcano--Gaussian using the Box--Muller method \citep{Ahn2002}. In general, there is a negligible impact on the emergent spectra for constant values $x_\text{crit} \ll (a\tau_0)^{1/3}$ or dynamical core skipping as in \citet{Smith2015}. However, for certain tests where the line separation decreases, such as the Ly$\alpha$ and deuterium cases, core skipping causes an unphysical bias towards one of the lines if they have different optical depths. Therefore, to show the accuracy of the analytical solutions, we turned off core skipping for such runs. To remedy this, one must generalise the dynamic core-skipping algorithm to use the minimum of $0.2\,(a\tau_0)_i^{1/3}$ for the line and the frequency offset placing the photon in the influence of nearby neighbouring lines, e.g. $k_{x,i} \lesssim k_x/2$, but we leave this for a future study.

\begin{figure}
    \centering
    \includegraphics[width=0.475\textwidth]{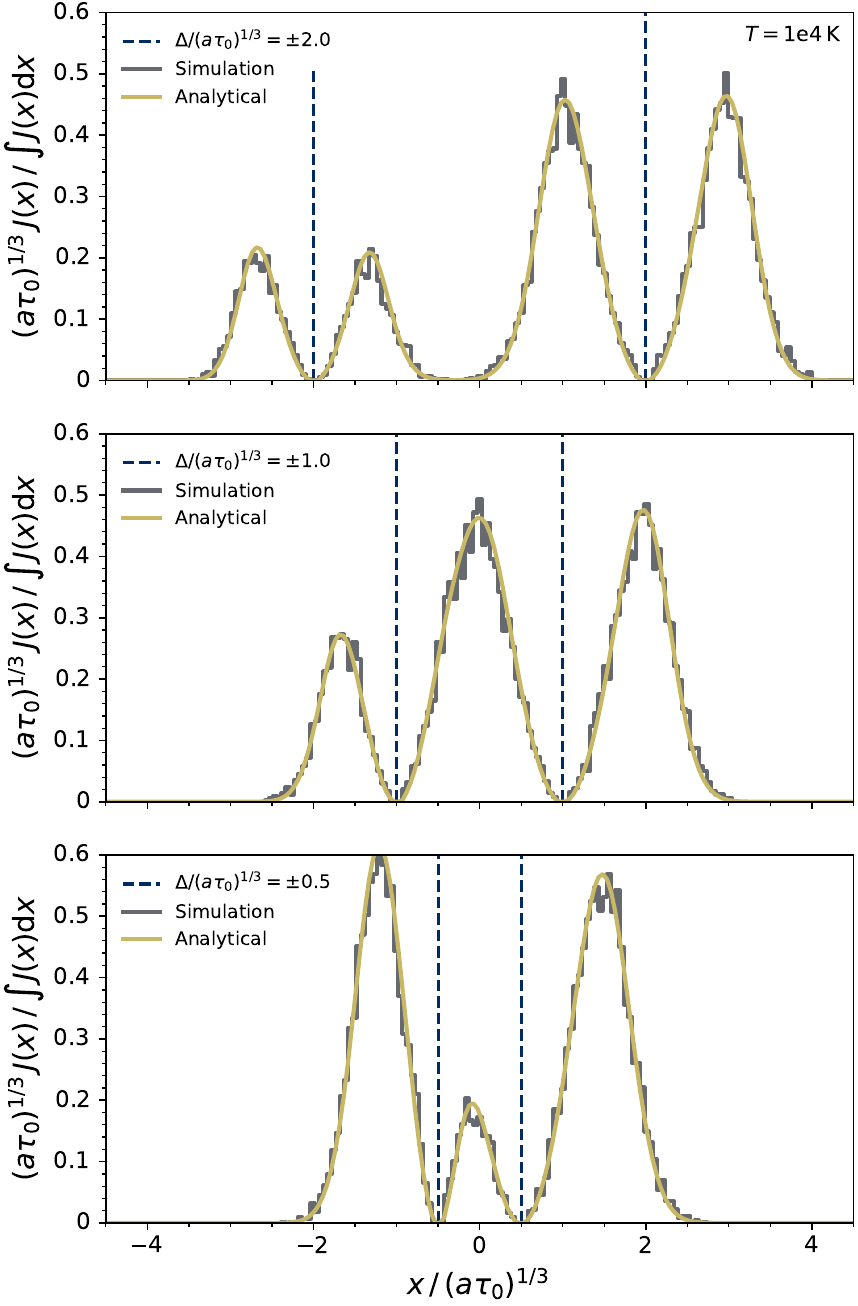}
    \caption{A toy network with two arbitrary lines of equal oscillator strength, illustrating the transition from well-separated lines with distinct peaks to overlapping lines with merged profiles as the frequency separation is reduced. We compare emergent spectra for slab geometry with different separations between the lines (top: widely separated; middle: intermediate; bottom: nearly overlapping). The analytic solution from Eq.~(\ref{eq:PS-emergent-solution-slab}) and numerical \textsc{colt} simulations are in excellent agreement even as the lines become highly coupled to produce nontrivial spectra.}
    \label{fig:general_tests}
\end{figure}

In Fig.~\ref{fig:general_tests}, we present results for the two-line V-shaped network with varying frequency separations between the lines, comparing these with our analytical solutions from Section~\ref{sec:equations}. As we decrease the line separation, the two distinct double-peaked solutions merge together like superimposed solutions. The physical reason for this is that once the line separation is $\Delta \lesssim (a\tau_0)^{1/3}$ the resonant broadening is enough to diffuse into the range of the other line, leading to mixing during random walk transport. We note that the flux approaches zero at each line centre because of the $\bar{H}(x)$ in the denominator has orders of magnitude higher cross-sections line centre, so frequency diffusion is required for escape. We also see that when the line separation becomes very small, the solution starts to recover the standard double-peaked structure of a singular line. The analytic solution is indistinguishable from the converged simulation output, confirming the validity of the Fokker--Planck approximation in high optical depth scenarios.

Finally, while we do have an analytical form of $\tilde{x} = \tilde{x}(x)$ as solved for in Appendix (\ref{ap:x tilde change of variables}), for this paper we converted between $\tilde{x}$ and $x$ space by numerically integrating Eq.~(\ref{eq:change_of_variables_true}). Despite the very strong agreement between the closed form solution and and numerical calculation, we chose to use the latter to avoid assumptions or approximations breaking down, and since it was computationally inexpensive.

\section{Applications to Real-World Systems} \label{sec:applications}
We now turn to two specific astrophysical applications to demonstrate V-shaped multi-line RT: Ly$\alpha$ fine-structure and deuterium injection. In both cases, multi-line effects are known to play a role, and our analytic solutions provide a convenient way to isolate these. The methods are also adaptable to other such systems.

\begin{figure}
    \centering
    \includegraphics[width=0.475\textwidth]{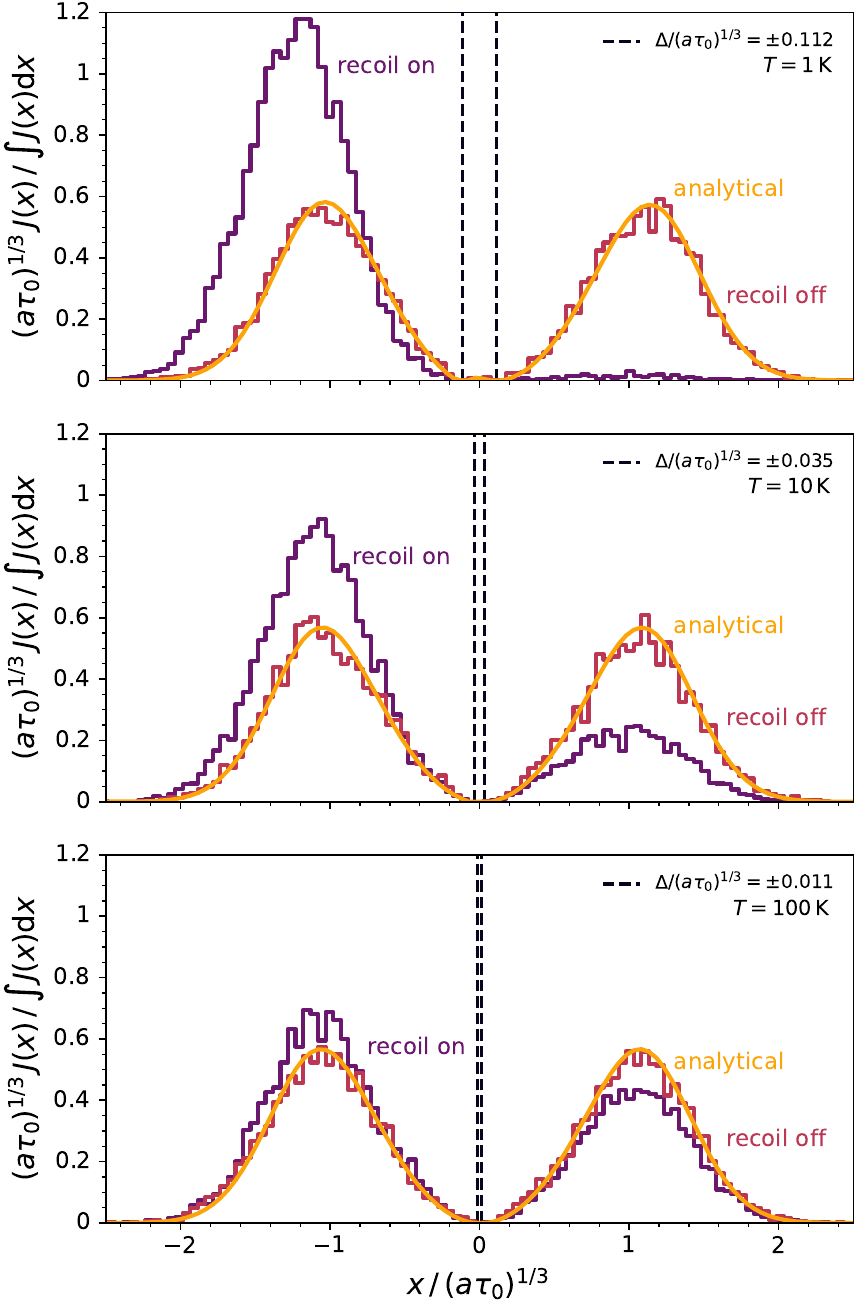}
    \caption{A physical Ly$\alpha$ fine-structure setup where we vary the temperature of the system. Even for very low gas temperatures the two lines do not cleanly separate in the emergent spectrum. We compare emergent Ly$\alpha$ spectra for spherical geometry at $T = \{1, 10, 100\}\,\text{K}$. Decreasing the temperature reduces the Doppler width and causes the line separation to increase. Even at $1$\,K, however, the two components overlap strongly and the profiles remain double-peaked rather than fully splitting. Atomic recoil becomes important at extremely low temperatures.}
    \label{fig:Lya}
\end{figure}

\subsection{Ly$\boldsymbol{\alpha}$ Fine Structure}

Relativistic corrections to the Schr\"{o}dinger equation and spin--orbit coupling causes the $2p$ energy level of the hydrogen atom to split into two distinct energy states. We ignore hyper-fine splitting of the ground state. This modifies the standard Ly$\alpha$ network, creating two closely spaced lines corresponding to the following atomic transitions: $1~{}^2S_{1/2} \longleftrightarrow 2~{}^2P^0_{1/2}$, and $1~{}^2S_{1/2} \longleftrightarrow 2~{}^2P^0_{3/2}$, which we denote as $1/2$ and $3/2$, respectively. The latter has twice the oscillator strength of the former, due to the $J$ degeneracy, but their wavelengths differ by $1.34\,\text{km\,s}^{-1}$. While fine-structure splitting is a more physically accurate picture of Ly$\alpha$ RT, most models ignore it because the separation becomes smaller than the Doppler broadening at gas temperatures higher than 100\,K \citep{SeonKim2020}. Since there is one lower state and two upper states, we can apply our solutions to this V-shaped two-line network. Inputting the relevant atomic data, we have $n_{1/2} = n_{3/2} = n_\text{H}$, and because both lines stem from hydrogen and $\nu_{1/2} \approx \nu_{3/2}$ we have $\Delta\nu_{\text{D}, 1/2} \approx \Delta\nu_{\text{D}, 3/2}$, such that $\mu_1 \approx \mu_2$; similarly since the Einstein coefficients are approximately equal this leads to damping parameters $a_{1/2} \approx a_{3/2}$. However, the weights differ according to $\omega_{1/2} \approx \frac{1}{2}\omega_{3/2} = 1/3$ due to the ratio of the oscillator strengths.

In Fig.~\ref{fig:Lya}, we present results from idealised cold clouds in slab geometry at three temperatures, $T = \{1, 10, 100\}\,\text{K}$, demonstrating excellent agreement between our analytic solution and numerical simulations. As temperature decreases, the thermal Doppler width decreases, exposing the fine-structure splitting of the Ly$\alpha$ line. However, even at 1\,K, the emergent spectra strongly resembles non-split Ly$\alpha$ because the peak locations are still much larger than the line separations $\Delta \ll (a\tau_0)^{1/3}$, blending the line interactions. It is important to note that our base solution becomes physically inaccurate once we consider the effect of recoil, as Ly$\alpha$ photons systematically lose energy to absorbing atoms. This would add a term to the redistribution function yielding \citep{Rybicki2006}
\begin{align}
    &\int J_{x'} H(x'_i)R_{x'-\Delta_i \rightarrow x-\Delta_i} \text{d}x' \nonumber \\
    &= H(x_i)J_{x} + \frac{1}{2}\frac{\partial}{\partial x}\left[H(x_i)\left( \frac{\partial J_{x}}{\partial x} +\frac{\Delta\nu_\text{D} hJ_x}{k_{\text{B}}T}\right)\right] \, .
\end{align}
Because of the inverse temperature term in the correction, recoil becomes more significant in cold gas, which matches the result of our simulations. To our knowledge, the only existing analytical solution including the effects of recoil is derived in \citet{Nebrin2025}, but this is only for a single line treatment and produces a series solution with coefficients defined recursively. We now outline how recoil might be incorporated into our multi-line framework. The full RTE becomes
\begin{align}
  &\frac{\bm{\nabla} }{k(\bm{r})} \bm{\cdot} \left(\frac{\bm{\nabla} J}{k(\bm{r})} \right) + \frac{3}{2}\sum_{i=0}^N\frac{\Delta\nu_{\text{D},i}^2}{\Delta\nu_\text{D}^2} H(x_i) \frac{\partial}{\partial x}\!\left[H(x_i)\!\left(\frac{\partial J}{\partial x} + \frac{h\Delta\nu_{\text{D},i}}{k_\text{B}T}J\right)\right] \nonumber \\
  &= \frac{3\bar{H}(x)}{k(\bm{r})}\int \frac{j_x}{4\pi} \, \text{d} \Omega \, .
\end{align}
We note that \citet{Nebrin2025} also includes other effects such as gas velocity gradients and continuum absorption/scattering, but as a proof of concept we neglect these effects and save them for a more comprehensive future study.

\begin{figure}
    \centering
    \includegraphics[width=0.475\textwidth]{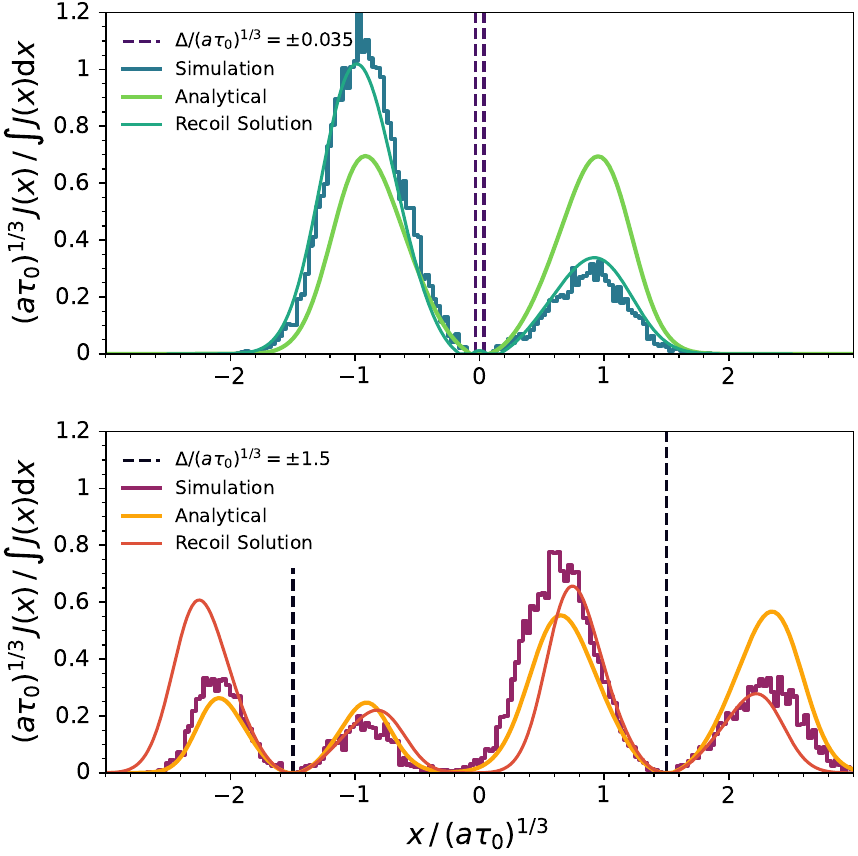}
    \caption{We compared the superimposed recoil solution with our non-recoil solution, along with \textsc{colt} simulations. Our high optical depth ($a\tau_0 = 10^5$), low temperature setup ($T = 10$\,K) ensures convergence of analytical solutions and strong recoil effects. \textit{Top:} Eq.~(\ref{eq:superimposed_recoil}) is used for fine-structure, and we see strong agreement. Note that the peaks are not perfectly aligned, an issue that is exacerbated for larger line separations. \textit{Bottom:} Eq.~(\ref{eq:superimposed_recoil}) is used for an arbitrary setup with larger line separation, and we observe misaligned peaks as well as incorrect heights, likely indicating that the line-specific parameters like optical depth or damping parameters are poorly modelled. In both panels, the qualitative features match, as we see the photons biased towards the red side of the plot, indicating that the true solution likely resembles this one with some non-linear correction.}
    \label{fig:recoil_plots}
\end{figure}

Although we do not incorporate recoil with full self-consistency, as an experiment in the regimes where a fine-structure treatment is needed, we combine our principle of superposition with the solution in \citet{Nebrin2025} to get
\begin{equation} \label{eq:superimposed_recoil}
    J(x) = \frac{\sqrt{18}\mathcal{L}}{(4\pi)^2\tau_0\Delta\nu_\text{D}R\bar{H}(x)} \sum_{n=1}^\infty \frac{2(-1)^n n\pi}{9(\mathcal{R}_+ - \mathcal{R}_-)} \sum_{i=0}^N \mathcal{F}\!\left(\frac{x-\Delta_i}{\beta_n}\right) \, ,
\end{equation}
where $\mathcal{R}_+, \mathcal{R}_-$, and $\mathcal{F}(x/\beta_n)$ are defined in appendix~A2 in \citet{Nebrin2025}. The results are illustrated in Fig.~\ref{fig:recoil_plots}, where we used the solutions for a fine-structure network (top) and a toy network with larger line separation (bottom) in the low temperature regime $T = 10$\,K.\footnote{At low temperatures the recoil solution converges slowly, so we used the Aitken delta-squared numerical acceleration scheme.} Naively applying the superposition yields results better than the single line case, but clearly the solutions are not sufficient, as evident in the large line separation case where the solutions produce the wrong peak locations and heights. A proper treatment would require generalising the single-line wing approximation $H(x,a) \approx a/\sqrt{\pi} x^2$ used in the derivation of this solution, and most likely require a double-wing or Wentzel--Kramers--Brillouin (WKB) approximation to obtain an analytically tractable differential equation. This is a formidable task with relatively little physical benefit, so we leave this for future studies.

In summary, our analytic solution is accurate and fine-structure splitting does not significantly alter emergent Ly$\alpha$ spectra unless the gas is extremely cold (although there may be other consequences for internal properties, such as the number of scatterings), at which point recoil must also be included.

\begin{figure}
    \centering
    \includegraphics[width=0.475\textwidth]{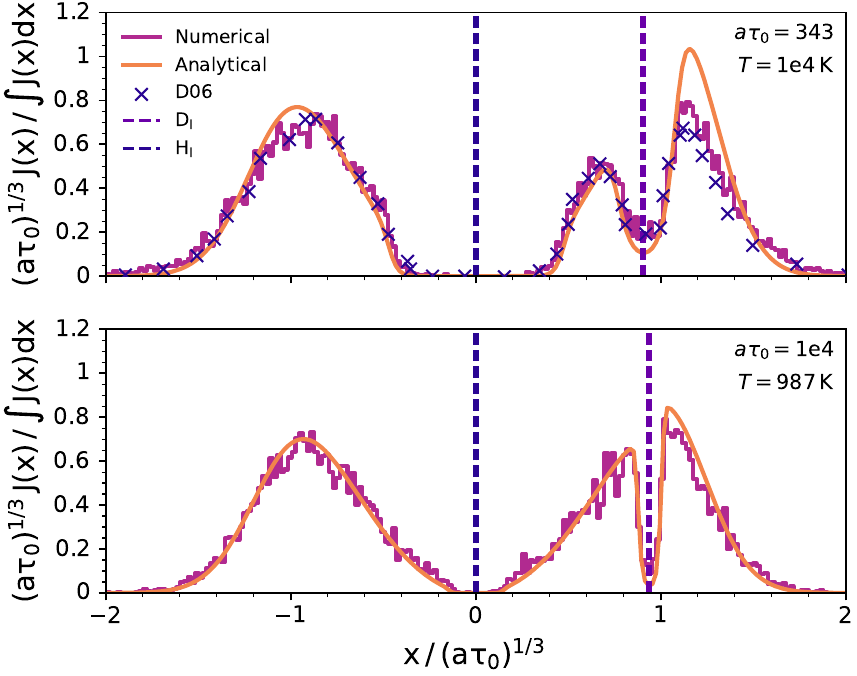}
    \caption{Emergent spectra of Ly$\alpha$ with deuterium injection, showing the analytical solutions give strong qualitative agreement with simulations, and converge in the regime $a\tau_0 \gtrsim 10^3$. The values of $a\tau_0$ and $T$ were specially chosen to align the deuterium absorption feature with the peak of the Ly$\alpha$ spectrum. \textit{Top:} We reproduce the setup in fig.~3 in \citet{Dijkstra2006} with $T = 10^4$\,K and moderate optical depth $a\tau_0 = 343$. The differences are due to the relatively low optical depth. \textit{Bottom:} With a higher optical depth $a\tau_0 = 10^4$ and lower temperature of $T = 987$\,K to align the absorption feature (which is also more narrow), we see much better agreement between our analytical solution and the emergent flux.}
    \label{fig:deuterium_spectra}
\end{figure}

\subsection{Deuterium Injection}

We also consider the effect that a small deuterium (D) fraction on Ly$\alpha$ RT. $\text{D}\,\textsc{i}$ has a Ly$\alpha$ transition nearly identical to that of hydrogen but slightly shifted in frequency due to the greater mass. Deuterium injection has been implemented into Monte Carlo codes \citep{Dijkstra2006, Michel-Dansac2020}, and while this is not exactly a V-shaped network since the lower energy levels are not the same energy, the lines are independent, so they can still be modelled with the same solution as before. Qualitatively, deuterium produces a narrow absorption dip on the blue side of the Ly$\alpha$ line. We use the ratio of deuterium atoms to hydrogen atoms as $n_\text{D}/n_\text{H} = 3.39 \times 10^{-5}$, obtained from \citet{BurlesTitler1998}. As described in \citet{Dijkstra2006}, the deuterium line centre frequency is shifted blueward by 82\,$\text{km\,s}^{-1}$. We choose to centre $x$ at Ly$\alpha$ line centre, so we have $x = (\nu - \nu_{0,\text{H}})/\Delta\nu_{\text{D},\text{H}}$. This corresponds to $\Delta_\text{H} = 0$ and $\Delta_\text{D} = +6.3\,T_4^{-1/2}$. Due to the doubled mass relative to hydrogen, factors of $\sqrt{2}$ must be accounted for, yielding $\mu_1 = 1$ and $\mu_2 = \sqrt{2}$. Deuterium, like hydrogen, will also encounter fine-structure splitting. This detail was ignored in previous analyses, but we have included it in our numerical and analytical implementation. Therefore, our network consists of four lines, with two fine structure Ly$\alpha$ lines each. Adopting the values of $\text{D}/\text{H}$, $f_{lu,i}$, and $\Delta\nu_{\text{D},i}$ for Ly$\alpha$ and deuterium into Eq.~(\ref{eq:H}), we get $\omega_{\text{H},1/2} = 0.333318$, $\omega_{\text{H},3/2} = 0.666634$, $\omega_{\text{D},1/2} = 1.59726 \times 10^{-5}$, and $\omega_{\text{D},3/2} = 3.1944\times 10^{-5}$. These weights indicate that the emergent spectra consists almost entirely of H Ly$\alpha$ emission.

In the top panel of Fig.~\ref{fig:deuterium_spectra}, we reproduce fig.~3 in \citet{Dijkstra2006}, and compare our analytical solution to the emergent spectra. The new solution qualitatively predicts the absorption feature caused by deuterium injection, albeit with a different peak height blueward of the deuterium line. However, $a_\text{H}\tau_{\text{H},0} = 343$, far less than the $a_\text{H}\tau_{\text{H},0} \gtrsim 10^3$ needed for analytical convergence of our solutions \citep{Neufeld1990, McClellan2022}. In the bottom panel of Fig.~\ref{fig:deuterium_spectra}, we increase the optical depth to $a_\text{H}\tau_{\text{H},0} = 10^4$ and lower the temperature to 987\,K, which improves convergence in both lines, since $a_\text{D}\tau_{\text{D},0}$ is also around 30 times higher than the previous test. However, we still have $a_\text{D}\tau_{\text{D},0} \ll 10^3$ so the deuterium line itself is not in the diffusion regime. Nevertheless, the overall agreement is good, and the analytic model captures the multi-line physics in sufficiently optically-thick media.

\begin{figure}
    \centering
    \includegraphics[width=0.475\textwidth]{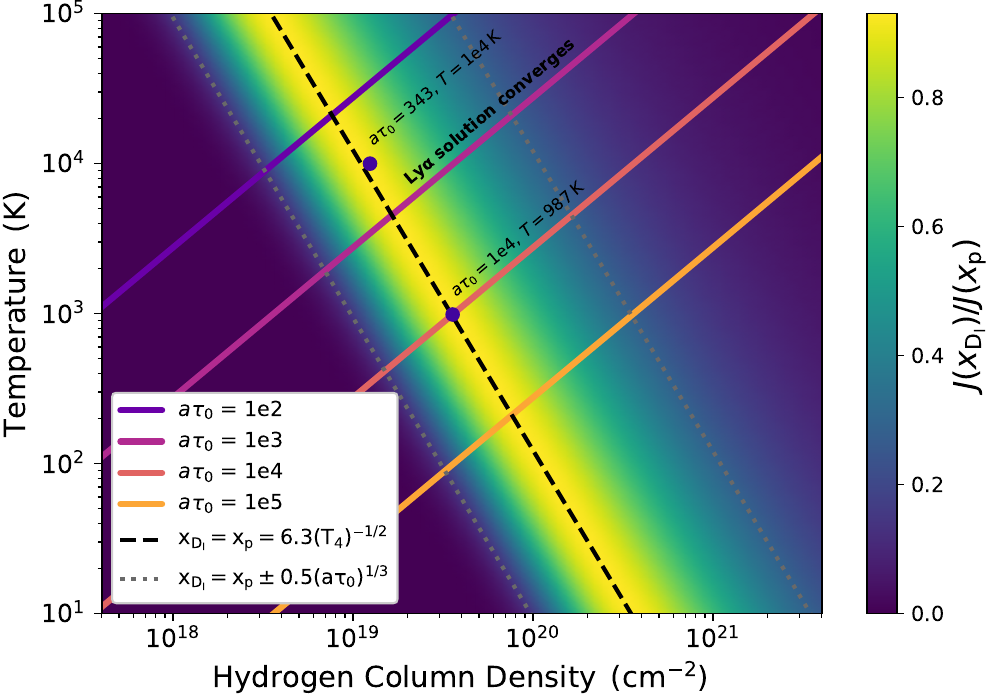}
    \caption{Deuterium absorption in column density and temperature parameter space. The black dotted line shows where the deuterium line core lies exactly on the peak of the Ly$\alpha$ flux, and the gray dotted lines indicate $\pm0.5\,(a\tau_0)^{1/3}$ away. The Ly$\alpha$ flux as a fraction of its peak is heat-mapped in the background.}
    \label{fig:validity_plot}
\end{figure}

\begin{figure}
    \centering
    \includegraphics[width=0.475\textwidth]{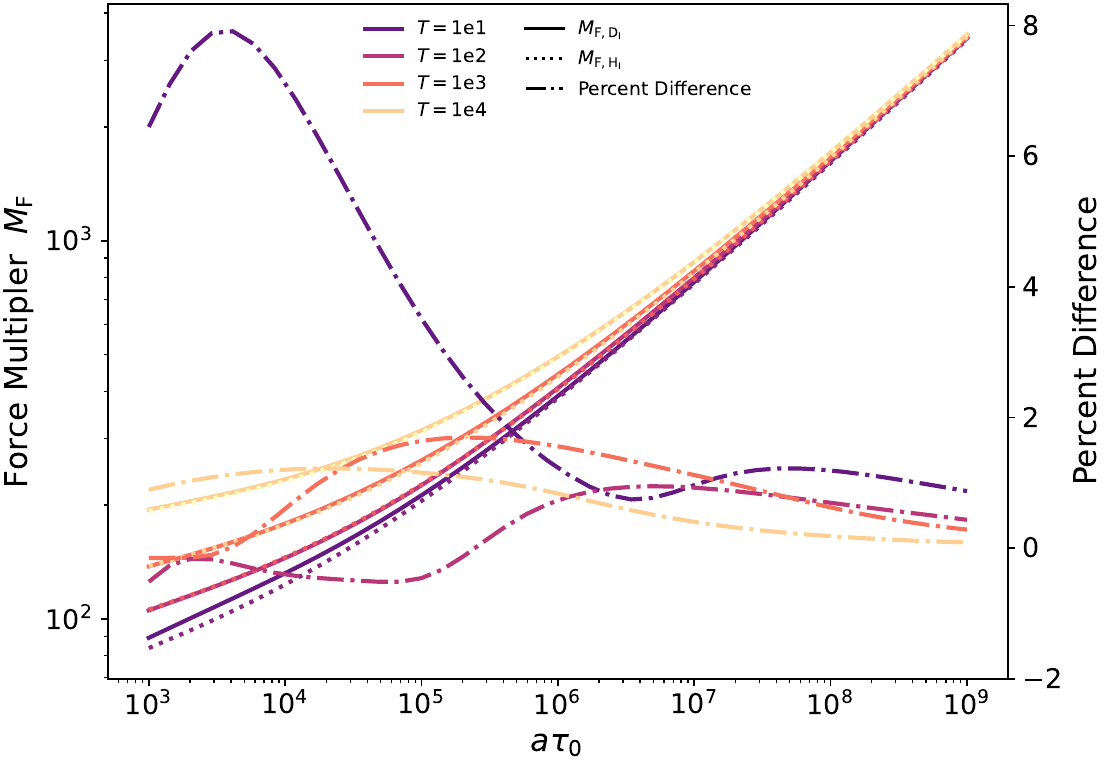}
    \caption{We numerically evaluate the force multiplier for hydrogen $(M_{\text{F},\,\text{H}})$ and for deuterium $(M_{\text{F},\,\text{D}})$ using Eq.~(\ref{eq:force_multipler}), and show that for most relevant astrophysical scenarios, the difference between the two is at the percent level. As temperature or optical depth increases, the force multipliers become equal.}
    \label{fig:force_multiplier}
\end{figure}

To determine when the deuterium line will have an observable impact in the Ly$\alpha$ spectrum, we demand that $x_\text{p} \approx 0.93099\,(a_\text{H}\tau_{\text{H},0})^{1/3} \approx \Delta_\text{D} = +6.3(T_4)^{-1/2}$, where $x_\text{p}$ is the peak of the Ly$\alpha$ solution \citep{LaoSmith2020}. Therefore, increasing the optical depth requires decreasing the temperature to align the absorption feature with the emergent flux. This relationship is shown in Fig.~\ref{fig:validity_plot}, where one can see which values of temperature ($T$) and hydrogen column density ($N_{\text{H}_\text{I}}$) cause the deuterium line centre to lie near the peak of Ly$\alpha$ solution. For a given optical depth, the alignment condition $\Delta_\text{D} = x_\text{p}$ yields $T = 2.77\times 10^7 \, (N_{\text{H}_\text{I}}/10^{20}\,\text{cm}^{-2})/(a\tau_0)$. Around this ``sweet spot'' region in parameter space, one must account for the effects of deuterium injection, while outside this range the absorption is either too close to the Ly$\alpha$ core or beyond the reach of resonant scattering. We caution that in realistic astrophysical environments, clouds have bulk motions (infall, outflow, and turbulence) that broaden and shift the Ly$\alpha$ line, likely smearing out deuterium absorption features across frequency space, making deuterium-induced signatures extremely challenging to observe. Our model of a static uniform cloud can be regarded as an optimistic upper limit on the impact of deuterium. Incorporating velocity gradients into the multi-line framework, building on \citet{Nebrin2025} and \citet{Smith2025}, would allow one to quantify the interplay of bulk motions and deuterium injection. We can also compare how deuterium affects the force multiplier with \citep{LaoSmith2020}:
\begin{align} \label{eq:force_multipler}
    M_{\text{F},\text{D}} &= \frac{-c}{3\mathcal{L}}\int \nabla u(r)\,\text{d}V = \frac{-c}{3\mathcal{L}}\int \frac{4\pi}{c}\int \nabla J(x) \,\text{d}x\text{d}V \\
    &= \frac{2\sqrt{6}}{3\pi}\sum_{n=1}^{\infty} \frac{[1-(-1)^n]}{n}\int \sum_{i=1}^4\left[\omega_i\mu_ie^{-\lambda_n |\tilde{x}-\tilde{\Delta}_i|} \right] \, \text{d}x(\text{d}\tilde{x}) \, . \notag
\end{align}
Unfortunately, the non-trivial change of variables means that our expression for $M_\text{F}$ is not closed form, but we can still compute this numerically. We also calculate the Ly$\alpha$ force multiplier using Eq.~(\ref{eq:force_multipler}) but with the single line treatment $N = 1, \,\bar{H}(x) \rightarrow H(x,a), \, \text{d}\tilde{x} = \sqrt{2/3}\,\text{d}x/[\tau_0H(x,a)]$ as in \citet{LaoSmith2020}. We show these quantities in Fig.~\ref{fig:force_multiplier}, and find that the difference is below two percent for most physical scenarios, except for low temperatures, where the low-$a\tau_0$ difference can be as high as eight percent. Therefore, deuterium injection does not change any prior conclusions about Ly$\alpha$ feedback, and another potential source of uncertainty in such modelling is removed \citep{Nebrin2025}.

\section{Discussion}
\label{sec:discussion}
We have developed and tested the first (to our knowledge) analytic framework for resonant RT involving multiple spectral lines simultaneously. Focusing on V-shaped atomic systems with one ground state and multiple excited states, i.e. coupled transitions, we derive closed-form solutions for the angle-averaged intensity in both infinite slab and spherical geometries. These solutions generalise previous single-line treatments \citep[e.g.][]{Harrington1973, Neufeld1990, LaoSmith2020, Nebrin2025} by showing that, under the same diffusion and high optical-depth approximations, the multi-line problem can be expressed as a weighted linear superposition of single-line solutions in an appropriately transformed frequency coordinate. This represents a powerful and somewhat unexpected conceptual advance: the complicated coupling between lines is encoded entirely in the composite absorption profile $\bar{H}$, frequency variable transformation $\tilde{x}(x)$, and weighting coefficients $\omega_i$ and $\mu_i$, making multi-line transfer analytically tractable for the first time.

Our comparisons with MCRT simulations reinforce the validity of the analytic solutions. By extending the \textsc{colt} code to handle multi-line scattering, we were able to simulate idealised and physical scenarios with overlapping lines and directly verify that the intensity profiles match the numerical ones to within the sampling noise. The agreement holds across a broad range of line separations, optical depths, and temperatures, as long as we remain in the regime where the diffusion approximation is valid, i.e. $a\tau_0 \gtrsim 10^3$ for each line. In cases where one line has a much lower optical depth than this, such as with trace deuterium injection, the solution can still predict qualitative effects but may not capture peak heights exactly. Still, considering the deuterium optical depths are often orders of magnitude below the expected criterion for applicability, we consider the multi-line solutions to be successful. This strong agreement confirms that the generalised absorption coefficient, modified Fokker--Planck approximation, and generalised change of variables accurately capture the underlying physics of resonant scattering and that the newly implemented multi-line extension to \textsc{colt} behaves as expected. Furthermore, the Monte Carlo method and diffusion solvers can now be applied to more complex problems beyond what analytic solutions can handle, providing a numerical laboratory for future studies.

The Ly$\alpha$ fine-structure application highlights both the strengths and limitations of the formalism. On one hand, the model correctly recovers the merging of the $2~{}^2P^0_{1/2}$ and $2~{}^2P^0_{3/2}$ components with increasing temperature. On the other hand, at very low temperatures the neglect of atomic recoil introduces a systematic discrepancy, since recoil adds an additional drift term in frequency space that becomes significant, producing asymmetric red and blue peaks when Doppler broadening is minimal ($T \lesssim 100$\,K). In an initial attempt to account for recoil, we also attempted to use the superposition principle on the solution of \citet{Nebrin2025}, turning their single-line results into a multi-line solution. However, the agreement with MCRT simulations is poor, especially as the line separation increases. Future work could involve relaxing the approximations used in the single-line derivation of \citet{Nebrin2025} to generate generalised solutions that include more complex effects like recoil, velocity gradients, or dust absorption. One could also consider incorporating polarisation, which may have interesting multi-line effects.

The deuterium injection case demonstrates the versatility of the approach for realistic astrophysical multi-species mixtures. Despite being outside the strict V-shaped configuration, the photons see two independent resonances, so the method successfully predicts the deuterium absorption feature and its evolution with optical depth, reproducing the qualitative behaviour seen in \citet{Dijkstra2006}. By surveying parameter space, we were able to identify where deuterium is most likely to matter along a narrow region of column density and temperature. The improved agreement at higher optical depths indicates that the analytic formalism converges to the exact solution in the same diffusion-dominated limit as single-line transfer. This opens the possibility of applying the model to other chemically or isotopically mixed media where multiple resonant transitions overlap, such as dense nebular regions, metal-enriched outflows, or partially ionised circumgalactic gas. We also find a negligible impact on the force multiplier, which means that recent results indicating the importance of Ly$\alpha$ radiation pressure feedback, but neglecting deuterium scattering, remain unchanged \citep[e.g.][]{Smith2017, Tomaselli2021,Nebrin2025,Manzoni2025}.\footnote{\cite{Kimm2018} did include deuterium scattering in their calculation of the force multiplier, estimated using the \textsc{rascas} MCRT code \citep{Michel-Dansac2020}. However, the authors did not isolate the effect of deuterium. Furthermore, their estimate of the force multiplier was later found to be incorrect as a result of numerical issues \citep[see discussion in][]{Nebrin2025}.} However, once again the model reaches its limits, as bulk velocity motions and turbulence can smear the deuterium absorption feature across frequency space, an effect not accounted for in this paper. Future studies could straightforwardly incorporate details from \citet{Nebrin2025} and \citet{Smith2015}. Each additional effect adds complexity, but our work provides a baseline solution upon which others can explore further.

Beyond the specific applications we examined, the present work provides the foundation for a more general theory of multi-line RT networks. The recent work of \citet{Chang2025}, who performed MCRT modelling of Rayleigh and Raman scattering, and branching transitions in between, underscores the importance of developing accurate multi-line RT methods. Their simulations show that multi-line scattering can produce non-intuitive line ratios and broadening that challenge simple interpretations of observed spectra. Extending our framework to include branching transitions would enable analytic exploration of systems where photons can scatter between multiple excited states before de-excitation, such as stellar outflows, accreting compact binaries, and Little Red Dots \citep{Harikane2023, Kocevski2023, Greene2024, Matthee2024, Chang2025}. For instance, $\Lambda$-shaped networks with two lower levels and one upper level or a cascade where an excited state can decay via multiple paths (as in more complete hydrogen treatments) would be natural next targets. Initial explorations are promising, but require a more careful treatment of generalised frequency redistribution functions. Bridging the gap between detailed numerical RT and analytic understanding will enable us to more effectively interpret the ever-increasing wealth of spectral data revealing unexpected phenomena. Such results establish a benchmark for future studies of multi-line resonant transfer and supply an analytic counterpart to numerical codes, facilitating rapid exploration of parameter space and deeper physical insight into the coupling of radiation and matter across multiple spectral lines.

\section*{Acknowledgments}

We thank the referee for constructive comments which have improved the quality of this work. We thank The University of Texas at Dallas Physics REU for bringing this collaboration together and providing various services throughout the research process. We also thank the National Science Foundation for funding through NSF grant PHY 2348872. All computations were performed on the Juno cluster supported by UT Dallas. AS acknowledges support through HST AR-17859, HST AR-17559, and JWST AR-08709. Finally, we thank the open source community of \textsc{NumPy} \citep{NumPy}, \textsc{SciPy} \citep{SciPy}, and \textsc{Matplotlib} \citep{Matplotlib} for providing extensive computational tools for data analysis in \textsc{Python}.

\bibliographystyle{mnras}

\bibliography{main}

\begin{appendix} 

\section{Solution to N-Line V-Shaped Networks} \label{ap:analytical solution}

We will apply the standard approximations used in other works \citep{Harrington1973, Neufeld1990, Dijkstra2006, LaoSmith2020, Nebrin2025, Lorinc2025} to simplify Eq.~(\ref{eq:RTE nu}). We first define angle-averaged moments of intensity as $J_x \equiv \frac{1}{4\pi} \int \text{d} \Omega \, I_x$ and $\bm{H}_x \equiv \frac{1}{4\pi} \int \text{d} \Omega \, I_x \bm{n}$. We use the bar to simply denote that this is the solution for a multi-line network. Using the definitions from Section~\ref{sec:equations}, the angle-averaged form of Eq.~(\ref{eq:RTE nu}) is:
\begin{align} \label{eq:zero-order-RT-eq}
  \frac{1}{c}\frac{\partial J_x}{\partial t} + \bm{\nabla} \bm{\cdot} \bm{H}_x &= \int \frac{j_x}{4\pi}\, \text{d} \Omega - k_x J_x \nonumber \\
  &+ k(\bm{r}) \sum_{i = 1}^N \omega_i \int J_{x'} H(x'_i, a_i) R_{x'-\Delta_i \rightarrow x - \Delta_i}\,\text{d}x' \, ,
\end{align}
where $R_{x' \rightarrow x} = \frac{1}{4\pi}\int \text{d}\Omega \, R_{x', n' \rightarrow x, n}$ is the angle-averaged redistribution function. We will assume steady state solutions and also assume a high optical depth environment, meaning that we can apply Fick's law to the moment equations combined with the Eddington approximation: 
\begin{align} \label{eq:ficks-law-insertion}
    &\frac{\partial J_x}{\partial t} = 0 \quad \text{and} \quad \bm{H}_x \approx -\frac{\nabla J_x}{3 k_x} \implies \\
    &\bm{\nabla} \bm{\cdot} \left(\frac{\bm{\nabla} J_x}{3 k_x} \right) = -\int \frac{j_x}{4\pi} \text{d} \Omega \, + k_x J_x - k(\bm{r}) \sum_{i = 1}^N \omega_i \int J_{x'} H(x'_i, a_i) R_{x'-\Delta_i \rightarrow x-\Delta_i}\,\text{d}x' \,.
\end{align}
To simplify the redistribution terms, start with the single-line frequency form of the Fokker--Planck approximation \citep{Rybicki1994}: 
\begin{equation}
  \int k_\nu R_{\nu' \rightarrow \nu} J_{\nu'}\,\text{d}\nu' \approx k_\nu J_\nu + \dfrac{1}{2} \Delta \nu_{\rm D}^2 \dfrac{\partial}{\partial \nu} \left( k_\nu \dfrac{\partial J_\nu}{\partial \nu} \right) \, .
\end{equation}
Transforming the derivatives to $x$ coordinates and using our notation for $k_x$:
\begin{align}
  &k(\bm{r})\omega_i\int H(x_i)R_{x'-\Delta_i \rightarrow x-\Delta_i} J_{x'}\,\text{d}x' \nonumber \\
  &\approx k(\bm{r})\omega_i \left[H(x_i) J_x + \dfrac{1}{2}\frac{\Delta \nu_{{\rm D},i}^2}{\Delta \nu_{{\rm D}}^2} \dfrac{\partial}{\partial x} \left( H(x_i) \dfrac{\partial J_x}{\partial x} \right) \right]
\end{align}
which when substituted into the redistribution integral yields
\begin{align} \label{eq:redistribution-approximation-insertion}
    &k(\bm{r}) \sum_{i = 1}^N \omega_j \int J_{x'} H(x'_i, a_i) R_{x'_i \rightarrow x_i}\,\text{d}x' \nonumber \\ &= k(\bm{r})\sum_{i=1}^N \omega_j \Bigg[H(x_i) J_{x} + \frac{1}{2\mu_i^2}\frac{\partial}{\partial x}\left(H(x_i) \frac{\partial J_{x}}{\partial x} \right) \, \Bigg] 
\end{align}
where $k_x$ is the same as defined in (\ref{eq:k_x}). 
We also define the functions $\tilde{H}$ and $\hat{H}$
\begin{equation}
    \tilde{H}(x) = \sum_{i=1}^{N}\omega_i\mu_iH(x_i, a_i) \quad 
    \text{and} \quad     
    \hat{H}(x) = \sum_{i=1}^{N}\frac{\omega_i}{\mu_i^2}H(x_i, a_i) \, ,
\end{equation}
which will become useful in the next few steps. Next, we split the luminosity source into spatial, frequency, and angular dependence with $j_x/(4\pi) = \mathcal{L}\eta(\bm{r})\frac{\tilde{H}(x)}{\sqrt{\pi}}\frac{1}{4\pi}$ such that $\iiint j_x/(4\pi) \,\text{d}x\text{d}V\text{d}\Omega = \mathcal{L}$ (note the $\tilde{H}$ for proper normalisation) , and substitute Eq.~(\ref{eq:redistribution-approximation-insertion}) into Eq.~(\ref{eq:ficks-law-insertion}). Dividing by $k(\bm{r})$ and multiplying by $3\bar{H}(x)$, we get 
\begin{equation} \label{eq:initial_A}
  \frac{1}{k(\bm{r})} \bm{\nabla}  \bm{\cdot} \left(\frac{\bm{\nabla} J}{k(\bm{r})} \right) + \frac{3}{2} \bar{H}(x) \frac{\partial}{\partial x} \left( \hat{H}(x) \frac{\partial J}{\partial x} \right) = -\frac{3\mathcal{L}}{4\pi} \frac{\eta(\bm{r})}{k(\bm{r})} \frac{\bar{H}\tilde{H}}{\sqrt{\pi}} \, .
\end{equation}
It is here that we make the following change of variables in position-space as was done in \citet{LaoSmith2020} as well as the transformation $\tilde{x} = \tilde{x}(x)$: 
\begin{equation} \label{eq:change-of-variables}
    \tilde{\bm{\nabla}} \equiv \frac{\tau_0}{k(\bm{r})} \bm{\nabla}
    \quad \implies \quad
    \tilde{\bm{r}}
    = \text{min} \int_0^r \frac{k(\bm{r'})}{\tau_0} d\bm{r}' \, , \quad \text{and} \quad \text{d}\tilde{x} = \sqrt{\frac{2}{3}}\frac{\text{d}x}{\tau_0 (\bar{H}\hat{H})^{1/2}} \, \, ,
\end{equation}
where $\tau_0 = k(\bm{r})L$ with $L$ being an arbitrary characteristic length of the system. To get $\tilde{x} = \tilde{x}(x)$ we direct the reader to Appendix~\ref{ap:x tilde change of variables}. We seek to simplify the right side of Eq.~(\ref{eq:initial_A}) and so in the spirit of \citet{LaoSmith2020} we note that $\bar{H}\hat{H}$ is sharply peaked at all $x_i$ like a series of delta functions, and also that $\int 3 \tau_0 \bar{H}(\tilde{x})\tilde{H}(\tilde{x}) \text{d}\tilde{x} = \int \sqrt{6} \tilde{H}\sqrt{\bar{H}/\hat{H}}\,\text{d}x \approx \sqrt{6 \pi}\sum\omega_i\mu_i$, which allows the substitution $3 \tau_0 \bar{H}\tilde{H}(\tilde{x}) \approx \sqrt{6\pi} \sum_{i=1}^N \omega_i\mu_i^2\delta(\tilde{x}_i)$ where $\tilde{x}_i = \tilde{x}(x_i)$. Finally, we let $J = \tilde{J} \mathcal{L} \tau_0 \sqrt{6} / (4\pi)$ we get a simplified, inhomogeneous Helmholtz equation: 
\begin{equation} \label{eq:simplified helmholtz}
    \tilde{\bm{\nabla}}^2 \tilde{J} + \frac{\partial^2 \tilde{J}}{\partial \tilde{x}^2} = - \frac{\eta(\tilde{\bm{r}})}{k(\tilde{\bm{r}})} \sum_{i=1}^{N} \omega_i\mu_i\,\delta(\tilde{x}-\tilde{\Delta}_i) \, .
\end{equation}
This equation differs from \citet{LaoSmith2020} only by the sum of delta functions and the coordinate transform $x \rightarrow \tilde{x}$.
The boundary conditions are the same as the single-line case, namely that
\begin{equation}
\label{eq:boundary conditions}
  \left[\tilde{\partial}_{\bm{s}} \tilde{J} + f \tau_0 \bar{H}(\tilde{x}) \tilde{J} \right]_{\bm{s}} = 0 \qquad \text{and} \qquad \lim_{\tilde{x} \rightarrow \pm \infty} \tilde{J} = 0 \, ,
\end{equation}
where $\bm{s}$ denotes the surface of the boundary and $\partial_{\bm{s}}$ denotes a derivative in the normal direction of the outward surface. This permits a Sturm--Liouville expansion \citep{Unno1955}, where we separate the eigenfunctions into spatial and frequency coordinates as 
\begin{equation} \label{eq:Sturm-Liouville expansion}
    \tilde{J}(\tilde{\bm{r}},\tilde{x}) = \sum_{n=1}^\infty \vartheta_n(\tilde{\bm{r}}) \varphi_n(\tilde{x}) \, ,
\end{equation}
which after plugging into Eq.~(\ref{eq:simplified helmholtz}) and using orthogonality yields our two characteristic equations:
\begin{align} \label{eq:characteristic equations}
    &\tilde{\nabla}^2 \vartheta_n + \lambda_n^2 \vartheta_n = 0 \, , \notag \\
    &\frac{\text{d}^2 \varphi_n}{\text{d}\tilde{x}^2} - \lambda_n^2 \varphi_n = -\frac{Q_n}{\tau_0} \sum_{i=1}^{N} \omega_i\mu_i^2 \delta(\tilde{x}_i)  = -\frac{Q_n}{\tau_0} \sum_{i=1}^{N} \omega_i\mu_i \delta(\tilde{x}-\tilde{\Delta}_i)\, .
\end{align}
where 
\begin{equation} \label{eq:general_Qn}
  Q_n = \tau_0\int \frac{\eta(\tilde{\bm{r}})}{k(\tilde{\bm{r}})} \vartheta_n(\tilde{\bm{r}})\,\text{d}\tilde{V} \, .
\end{equation}
Since $\vartheta_n$ and $Q_n$ depends on the geometric setup, we will save them for the next sections and solve for $\varphi_n$. Since our applications will be for two lines, we will set $N=2$ and solve Eq.~(\ref{eq:characteristic equations}) with $\lim_{\tilde{x}\rightarrow\pm\infty} \tilde{J} = 0$, and the jump conditions $\Delta (\text{d}\varphi_n/\text{d}\tilde{x})_{\tilde{x} = \Delta_1} = -Q_n \omega_1 \mu_1 / \tau_0$ and $\Delta (\text{d}\varphi_n/\text{d}\tilde{x})_{\tilde{x} = \Delta_2} = -Q_n \omega_2 \mu_2 / \tau_0$ obtained from integration around the boundaries, where we let $\Delta_1 < \Delta_2$. The solution is 
\begin{equation}
    \varphi_n = \begin{cases}
        Ae^{\lambda_{n}\tilde{x}} & \tilde{x} < \tilde{\Delta}_1 \\
        Be^{\lambda_{n}\tilde{x}} + Ce^{-\lambda_{n}\tilde{x}} & \tilde{\Delta}_1 < \tilde{x} < \tilde{\Delta}_2 \\
        De^{-\lambda_{n}\tilde{x}} & \tilde{\Delta}_2 < \tilde{x}
    \end{cases}
\end{equation}
where 
\begin{equation}
    \begin{cases}
    A = \frac{Q_{n}}{2\tau_{0}\lambda_{n}} ( \omega_1 \mu_1 e^{-\lambda_{n}\tilde{\Delta}_1} + \omega_2 \mu_2 e^{-\lambda_{n}\tilde{\Delta}_2} ) \\
    B = \frac{Q_{n}\omega_2 \mu_2}{2\tau_{0}\lambda_{n}}e^{-\lambda_{n}\tilde{\Delta}_2} \\ 
    C = \frac{Q_{n}\omega_1 \mu_1}{2\tau_{0}\lambda_{n}}e^{\lambda_{n}\tilde{\Delta}_1} \\ 
    D = \frac{Q_{n}}{2\tau_{0}\lambda_{n}} ( \omega_1 \mu_1 e^{\lambda_{n}\tilde{\Delta}_1} + \omega_2 \mu_2 e^{\lambda_{n}\tilde{\Delta}_2} ) \, .
\end{cases} 
\end{equation} 
Putting this all together and reinserting into Eq.~\ref{eq:Sturm-Liouville expansion} we have a solution of
\begin{equation}
  J(\tilde{\bm{r}},\tilde{x}) = \frac{\mathcal{L} \sqrt{6}}{8 \pi} \sum_{n=1}^\infty \frac{Q_{n}}{\lambda_{n}} \vartheta_n(\bm{r}) \left( \omega_1 \mu_1 e^{-\lambda_{n}|\tilde{x}-\tilde{\Delta}_1|} + \omega_2 \mu_2 e^{-\lambda_{n}|\tilde{x}-\tilde{\Delta}_2|} \right) \, .
\end{equation}
This is a weighted superposition of single-line solutions, which completes the derivation of Eq.~(\ref{eq:superposition1}). For other solutions, we adopt the results from \citet{LaoSmith2020} to solve for a specific geometrical setup. 
Our infinite slab is as described in Section~\ref{sec:analytics} with $k(z)$ and finite thickness $\tau_0$ in the $\pm z$ direction for total optical depth $2\tau_0$. For physical solutions, we add another condition where we only seek even solutions about the $xy$ plane. Eq.~(\ref{eq:change-of-variables}) then becomes
\begin{equation}
    \tilde{\bm{r}}
    = \text{min} \int_0^r \frac{k(\bm{r'})}{\tau_0} d\bm{r}' \quad \longrightarrow \quad 
    \tilde{z} = \int_0^z\frac{k(z')}{\tau_0}\text{d}z' \,.
\end{equation}
The eigenfunctions have already been solved, and the solution is 
\begin{equation} \label{eq:eigenfunctions-slab}
    \vartheta_n = \text{cos}(\lambda_n \tilde{z}) \quad \text{where} \quad n= 1, 2, 3 \dotsc 
\end{equation}
and the eigenvalues are solved from the boundary conditions which generate this transcendental equation 
\begin{equation} \label{eq:transcendental-slab}
    \lambda_n\text{tan}(\lambda_n) = f\tau_0\bar{H}(\tilde{x}) \, ,
\end{equation}
with the zeroth order eigenvalues being \citep{Harrington1973}
\begin{equation} \label{eigenvalues-slab}
    \lambda_n = \pi\left(n-1 \right) + \text{tan}^{-1}\left(\frac{f\tau_0\bar{H}(\tilde{x})}{\lambda_n} \right)  \approx \pi \left(n-\frac{1}{2} \right) \, .
\end{equation}
For a point source, the radial component of the emission coefficient becomes $\eta(z) = \delta(z)$, which leads to $Q_n = 1$. Since we want a point source for the emission, we let $\eta(z) = \delta(z)$ and so 
\begin{equation}
\label{eq:Q_n-slab}
  Q_n = 2 \tau_0 \int_0^1 \frac{\eta(\tilde{z})}{k(\tilde{z})}\text{cos}(\lambda_n\tilde{z})\text{d}\tilde{z} = 1 \, .
\end{equation}
Plugging this all in and summing, we get $J(\tilde{z},\tilde{x})$ from Eq.~(\ref{eq:PS-interior-solution-slab}). To obtain the emergent spectra, we use Eqs.~(\ref{eq:transcendental-slab}) and (\ref{eigenvalues-slab}) to approximate 
\begin{equation}
    \text{cos}(\lambda_n) \approx \frac{\lambda_n \text{sin}(\lambda_n)}{f\tau_0\bar{H}(\tilde{x})} \approx \frac{\lambda_n (-1)^{n-1}}{f\tau_0\bar{H}(\tilde{x})} \, ,
\end{equation}
which gives our solution at the boundary to be $J(\tilde{x})$ from Eq.~(\ref{eq:PS-emergent-solution-slab}). The spherical geometry setup is a perfect, idealised sphere with constant absorption $k_0$ and thickness $R$ corresponding to optical depth $\tau_0 = k_0 R$.
Eq.~(\ref{eq:change-of-variables}) then becomes
\begin{equation}
    \tilde{\bm{r}}
    = \text{min} \int_0^r \frac{k(\bm{r'})}{\tau_0} d\bm{r}' \quad \longrightarrow \quad 
    \tilde{r} = \frac{k_0}{\tau_0} r \, ,
\end{equation}
where $r$ is of course radial distance. For physical solutions, we add another boundary condition where the solution must not diverge at $r = 0$. Once again, the eigenfunctions are already known \citep{Dijkstra2006} and are
\begin{equation} \label{eq:eigenfunctions-sphere}
  \vartheta_n = \frac{1}{\sqrt{2\pi}}\frac{\sin(\lambda_n \tilde{r})}{\tilde{r}} \qquad \text{where} \quad n = 1, 2, \ldots \, ,
\end{equation}
and the eigenvalues are solved from the boundary conditions which generate this transcendental equation
\begin{equation} \label{eq:transcendental-sphere}
  \lambda_n \cot(\lambda_n) = 1 - f \tau_0 \bar{H}(\tilde{x}) \approx -f \tau_0 \bar{H}(\tilde{x}) \, ,
\end{equation}
with the zeroth order eigenvalues being
\begin{equation} \label{eq:eigenvalues-sphere}
  \lambda_n = \pi n + \tan^{-1}\left(\frac{\lambda_n}{1 - f \tau_0 \bar{H}(\tilde{x})}\right) \approx \pi n \, .
\end{equation}
Similarly, we choose a delta function emission source so let $\eta(r) = \delta(r)/(4\pi r^2) = \delta(\tilde{r})/(4 \pi \tilde{r}^2R^3)$ and so
\begin{equation} \label{eq:Qn-sphere}
  Q_n = \frac{4\pi R}{\sqrt{2\pi}} \int_0^1 \tilde{r} \eta(\tilde{r}) \sin(\lambda_n \tilde{r})\,\text{d}\tilde{r} = \lambda_n/(\sqrt{2\pi}R^2)\, .
\end{equation}
The interior solution then becomes $J(\tilde{r},\tilde{x})$ from Eq.~(\ref{eq:PS-interior-solution-sphere}). We again use a similar approximation on Eqs.~(\ref{eq:transcendental-sphere}) and (\ref{eq:eigenvalues-sphere}) to obtain
\begin{equation}
    \sin(\lambda_n) \approx (-1)^{n-1} \lambda_n/ f \tau_0 \bar{H}(\tilde{x})
\end{equation}
which is then used to get the solution on the boundary for $J(\tilde{x})$ from Eq.~(\ref{eq:PS-emergent-solution-sphere}). As shown in Appendix (\ref{ap:x tilde change of variables}), it is tedious to substitute $\tilde{x} \rightarrow x$ into our final solutions, and so we leave this solution and the ones to follow in $\tilde{x}$ space. 

\section{Change of Variables in Frequency Space}
\label{ap:x tilde change of variables}

Recall from Appendix~\ref{ap:analytical solution} that we used the Fokker--Planck approximation to simplify the redistribution term and cancel out the absorption term, yielding the following equation:
\begin{equation} \label{eq:initial}
    \frac{1}{k(\bm{r})} \bm{\nabla}  \bm{\cdot} \left(\frac{\bm{\nabla} J}{k(\bm{r})} \right) + \frac{3}{2} \bar{H}(x) \frac{\partial}{\partial x} \left( \hat{H}(x) \frac{\partial J}{\partial x} \right) = -\frac{3\mathcal{L}}{4\pi} \frac{\eta(\bm{r})}{k(\bm{r})} \frac{\bar{H}\tilde{H}}{\sqrt{\pi}} \, ,
\end{equation}
where 
\begin{equation}
    \bar{H}(x) = \sum_{i=1}^{N}\omega_iH(x_i,a_i) \,, \, \, \,      
    \hat{H}(x) = \sum_{i=1}^{N}\frac{\omega_i}{\mu_i^2}H(x_i,a_i) \,, \, \, \,     \tilde{H}(x) = \sum_{i=1}^N\omega_i\mu_iH(x_i,a_i) \, ,
\end{equation}
as before. We seek a change of variables $\tilde{x} = \tilde{x}(x)$ that satisfies (constants omitted)
\begin{equation} \label{eq:cov_want}
     \frac{\partial^2J_x}{\partial\tilde{x}^2} =\bar{H} \dfrac{\partial}{\partial x} \left( \hat{H} \dfrac{\partial J_x}{\partial x} \right) = \bar{H}\left[\left(\frac{\partial \hat{H}}{\partial x} \right)\left(\frac{\partial J_x}{\partial x} \right)  + \hat{H}\frac{\partial^2 J_x}{\partial x^2} \right] \, ,
\end{equation}
which will allow us to cleanly use Sturm--Liouville as in \citet{LaoSmith2020}.
Using the chain rule, we have that 
\begin{align} \label{eq:chain_rule}
    &\frac{\partial}{\partial \tilde{x}} = \frac{\partial x}{\partial \tilde{x}}\frac{\partial}{\partial x}\, , \nonumber \\
    &\frac{\partial^2}{\partial \tilde{x}^2} = \left(\frac{\partial x}{\partial \tilde{x}}\frac{\partial}{\partial x}\right) \left(\frac{\partial x}{\partial \tilde{x}}\frac{\partial}{\partial x}\right) = \frac{\partial x}{\partial \tilde{x}}\left[ \left(\frac{\partial^2 x}{\partial x\partial\tilde{x}}\right)\frac{\partial}{\partial x} + \frac{\partial x}{\partial \tilde{x}}\frac{\partial^2}{\partial x^2} \right] \, .
\end{align}
Matching with Eq.~(\ref{eq:cov_want}) and rearranging we have 
\begin{equation} \label{eq:cov_conditions}
    \left(\frac{\partial x}{\partial \tilde{x}} \right) = \sqrt{\bar{H}\hat{H}} \quad \text{and} \quad \frac{\text{d}}{\text{d}x}\text{ln}(\hat{H}(x)) = \frac{\text{d}}{\text{d}x}\text{ln}(\bar{H}(x)) \, .
\end{equation}
The first equation gives our change of variables, while the second is its convergence condition. The second equation is only satisfied for $\bar{H} = c\hat{H}$ for some constant $c$. However, even if $\bar{H} \neq c\hat{H}$ for all $x$, for when $x$ is close to line $i$ but far from the others, or $x_i \ll x_j$ for all $i \neq j$, then $\bar{H} \approx \omega_1H(x_1)$, and $\hat{H} \approx \omega_1/\mu_1H(x_1)$ where $\mu_i = \Delta\nu_\text{D}/\Delta\nu_{\text{D},i}$. Therefore, $\bar{H} \approx \mu_1\hat{H}$. In other words, Eq.~(\ref{eq:cov_conditions}) still holds quite well in the case where the lines are far apart. The difference can be computed as simply subtracting the single derivative term in Eq.~(\ref{eq:cov_want}) from Eq.~(\ref{eq:chain_rule}) to get 
\begin{align}
    \text{difference} &= \left(\frac{\partial x}{\partial \tilde{x}} \right)\frac{\partial}{\partial x}\left( \sqrt{\bar{H}\hat{H}} \right) - \bar{H}\frac{\partial \hat{H}}{\partial x} = \frac{1}{2}\left(\hat{H}\frac{\partial\bar{H}}{\partial x} + \bar{H}\frac{\partial\hat{H}}{\partial x} \right) - \bar{H}\frac{\partial \hat{H}}{\partial x} \nonumber \\
    & = \frac{1}{2}\left(\hat{H}\frac{\partial\bar{H}}{\partial x} - \bar{H}\frac{\partial\hat{H}}{\partial x} \right) \, ,
\end{align}
which becomes negligible as one line dominates. This approximation is good enough for our applications, and so our final change of variables is\footnote{We also note that the same basic result, i.e. a dependence on $\tilde{x}$, can be derived from a WKB approximate solution for the frequency eigenfunctions $\varphi_n$.}
\begin{equation} \label{eq:change_of_variables_true}
    \text{d}\tilde{x} = \sqrt{\frac{2}{3}} \frac{\text{d}x}{\tau_0(\bar{H}\hat{H})^{1/2}} \quad \implies \quad \tilde{x} = \int_0^{x}\sqrt{\frac{2}{3}}\frac{\text{d}x'}{\tau_0(\bar{H}\hat{H})^{1/2}} \, , 
\end{equation}
where we added the constant $(\sqrt{2/3})/\tau_0$ to help simplify Eq.~(\ref{eq:initial}). This is the form we use, and we numerically evaluate it for every figure in this paper.

Computing this integral analytically is difficult, so we approximate $\sqrt{\bar{H}\hat{H}} \approx \sum (\omega_i/\mu_i)H(x_i, a_i)$ which is justified in the case of large line separation. We also limit our solutions to the two line case. To proceed further, we will need to approximate the Voigt profile itself, which brings us to the two following cases:

\textbf{Wing + Wing Approximation:}
We recall Eq.~(\ref{eq:H}) and we approximate $H(x, a) \approx a/(\sqrt{\pi}x^2)$ for both lines, which is valid for $x_i \gtrsim x_{\text{cw},i} \sim 3$ \citep{Smith2015}. Therefore, we have
\begin{align}
    \tilde{x} &= \int_0^{x}\sqrt{\frac{2}{3}}\frac{\text{d}x'}{\tau_0\bar{H}(x')} \approx \sqrt{\frac{2\pi}{3}}\frac{1}{\tau_0}\int_0^{x}\frac{\text{d}x'}{\frac{\omega_1a_1}{\mu_1^3(x'-\Delta_1)^2} + \frac{\omega_2a_2}{\mu_2^3(x'-\Delta_2)^2}} \nonumber \\
    &= 
    \sqrt{\frac{2\pi}{3}}\frac{1}{\tau_0}\int_0^{x}\frac{\text{d}x'(x'-\Delta_1)^2(x'-\Delta_2)^2}{b_1(x'-\Delta_2)^2 + b_2(x'-\Delta_1)^2} \, ,
\end{align}
where $b_i = \omega_i a_i/\mu_i^3$. Because of the definition of $x$, $\Delta_1 = -\Delta_2$ and without loss of generality let $\Delta_2 > 0$. Solving this integral is lengthy but doable, and the result is 
\begin{equation} \label{eq:wing-wing}
    \tilde{x} \approx \sqrt{\frac{2\pi}{3}}\frac{1}{\tau_0}
    \Bigg[Ax'^3 + Bx'^2 + Cx' + D\text{ln}(E(x')) + F\text{arctan}\left(G(x')\right)
    \Bigg] \Bigg|_0^x \, ,
\end{equation}
where we define $\alpha = b_1 + b_2$ and $\beta = b_2 - b_1$, and the other terms as $A = 1/(3\alpha)$, $B = -\Delta\beta/\alpha^2$, $C = \Delta^2(b_1^2 - 14b_1b_2 + b_2^2) / \alpha^3$, $D = 16 b_1 b_2 \beta \Delta^3 / \alpha^4$, $E(x) = x^2\alpha + 2x\Delta\beta + \Delta^2\alpha$, $F = -8b_1b_2(b_2^2 - 6b_1b_2 + b_1^2)\Delta^3 / (\sqrt{b_1b_2}\alpha^4)$, and $G(x) = (\alpha x + \Delta\beta^2) / (2\Delta\sqrt{b_1b_2})$. We illustrate this in Fig.~(\ref{fig:wing_wing_combined}).

\setcounter{figure}{0}
\renewcommand{\thefigure}{A\arabic{figure}}

\begin{figure}
    \centering
    \includegraphics[width=0.475\textwidth]{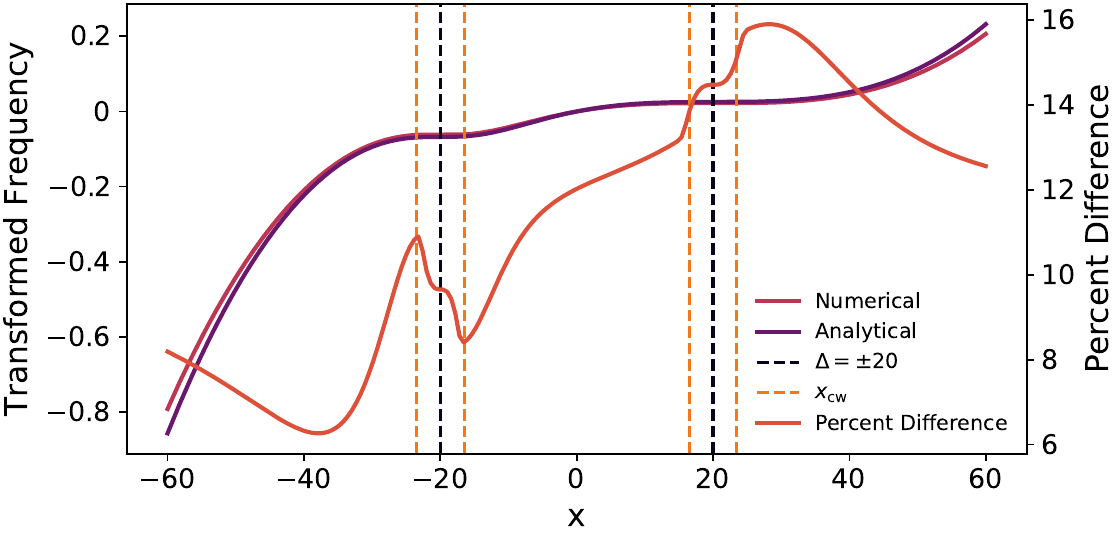}
    \caption{The analytical expression from Eq.~(\ref{eq:wing-wing}) shows strong agreement with numerical evaluation of Eq.~(\ref{eq:change_of_variables_true}), even in the core where the wing approximation breaks down. The error is asymmetric due to the asymmetric line parameters (see Appendix \ref{ap:parameter_table}).}
    \label{fig:wing_wing_combined}
\end{figure}

\textbf{Core + Wing Approximation:}
Another approximation worth using is using one line with the wing approximation and one with the core approximation, so without loss of generality let $H_1 = e^{-\mu_1^2(x+\Delta)^2}$ and $H_2 = a_2/(\sqrt{\pi}\mu_2^2(x-\Delta)^2)$. This leads to 
\begin{align}
    \tilde{x} &= \int_0^{x}\sqrt{\frac{2}{3}}\frac{\text{d}x'}{\tau_0\bar{H}(x')} \approx \sqrt{\frac{2}{3}}\frac{1}{\tau_0}\int_0^{x}\frac{\text{d}x'}{\omega_1 e^{-\mu_1^2(x'+\Delta_1)^2} + \frac{\omega_2a_2}{\sqrt{\pi}\mu_2^2(x'-\Delta_2)^2}} \nonumber \\
    &= 
    \sqrt{\frac{2}{3}}\frac{1}{\tau_0}\int_0^{x}\frac{\text{d}x'(x'-\Delta)^2}{b_1(x'-\Delta)^2 e^{-\mu_1^2(x'+\Delta)^2} + b_2} \, ,
\end{align}
where $b_1 = \omega_1$ and $b_2 = \omega_2a_2/(\sqrt{\pi}\mu_2^2)$. Unfortunately, this expression is not integrable, so we make further approximations. For $b_2 \ll b_1(x-\Delta)^2 e^{-(x+\Delta)^2}$, we simply get 
\begin{equation}  \label{eq:core-wing-small-b2}
    \tilde{x} = \sqrt{\frac{\pi}{6}}\frac{1}{b_1\mu_1\tau_0}\text{erfi}\left[\mu_1(x'+\Delta)\right]\Big|_0^x \,.
\end{equation}
For $b_2 \gg b_1(x'-\Delta)^2 e^{-\mu_1^2(x'+\Delta)^2}$, we take a first order approximation to get 
\begin{align}
    \tilde{x} &= \sqrt{\frac{2}{3}}\frac{1}{b_2\tau_0}\int_0^{x}\frac{\text{d}x'(x'-\Delta)^2}{\frac{b_1}{b_2}(x'-\Delta)^2 e^{-\mu_1^2(x'+\Delta)^2} + 1} \nonumber \\ 
    &=
    \sqrt{\frac{2}{3}}\frac{1}{b_2\tau_0}\int_0^{x}\text{d}x'(x'-\Delta)^2\Bigg[1 - \frac{b_1}{b_2}(x'-\Delta)^2 e^{-\mu_1^2(x'+\Delta)^2} + \dots \Bigg] \nonumber \\
    &\approx \sqrt{\frac{2}{3}}\frac{1}{b_2\tau_0}\Bigg[\frac{1}{3}(u- 2\Delta)^3 - \frac{b_1}{b_2}\Bigg[\frac{1}{\mu_1^4}\left(\frac{3\sqrt{\pi}}{8}\text{erf}(u) - \frac{1}{2}u^3e^{-u^2} - \frac{3}{4}ue^{-u^2}\right) \nonumber \\
    &- \frac{8\Delta}{\mu_1^3}\left(\frac{\sqrt{\pi}}{4}\text{erf}(u)- \frac{1}{2}u^2 e^{-u^2}\right) 
    + \frac{24\Delta^2}{\mu_1^2}\left(\frac{\sqrt{\pi}}{4}\text{erf}(u)-\frac{1}{2}ue^{-u^2}\right) \nonumber \\
    &+ \frac{16\Delta^3}{\mu_1}e^{-u^2} + 8\sqrt{\pi}\Delta^4\text{erf}(u)\Bigg] \Bigg] \Bigg|_0^{x}
\end{align}
where we used $u = \mu_1(x' + \Delta)$ to simplify the expression.

\section{Figure Parameters} \label{ap:parameter_table}
For reproducibility, below is a table of all parameters used throughout this paper to generate each figure.

\begin{table}[h]
\centering
\renewcommand{\arraystretch}{1.25}
{\setlength{\tabcolsep}{4pt}
\begin{tabular}{|c|c|c|c|c|c|c|c|c|}
\hline
\textbf{Fig.} 
& $a\tau_0$ 
& $T\,(\mathrm{K})$ 
& $a_1$ 
& $a_2$ 
& $\omega_1$ 
& $\omega_2$ 
& $\mu_1$ 
& $\mu_2$
\\
\hline

\ref{fig:H_bar}
& 1e5 & -- & 1e-3 & 1e-2 & 0.3 & 0.7 & 10 & 1  \\

\ref{fig:internal}a
& -- & -- & 1e-4 & 1e-3 & 1/4 & 3/4 & -- & --  \\

\ref{fig:internal}b
& 1e2 & -- & 1e-4 & 1e-3 & 1/4 & 3/4 & -- & -- \\

\ref{fig:J_2_individual}
& 1e2 & -- & 1e-4 & 1e-3 & 1/4 & 3/4 & 1 & 1  \\ 

\ref{fig:general_tests}
& 1e5 & 1e4 & 4.71e-4 & 4.71e-4 & 1/3 & 2/3 & 1 & 1  \\ 

\ref{fig:Lya}
& 1e5 & -- & 4.71e-4 & 4.71e-4 & 1/3 & 2/3 & 1 & 1  \\

\ref{fig:recoil_plots}
& 1e5 & 1e1 & 4.71e-4 & 4.71e-4 & 1/3 & 2/3 & 1 & 1\\ 

\ref{fig:deuterium_spectra}
& -- & -- & 4.71e-4 & 6.67e-4 & 0.999952 & 4.79e-5 & 1 & $\sqrt{2}$  \\ 

\ref{fig:validity_plot}
& -- & -- & 4.71e-4 & 4.71e-4 & 1/3 & 2/3 & 1 & 1  \\ 

\ref{fig:force_multiplier}
& -- & -- & -- & -- & -- & -- & -- & -- \\ 

\ref{fig:wing_wing_combined}
& 1e5 & -- & 1e-4 & 1e-4 & 1/4 & 3/4 & 1.2 & 0.8 \\ \hline
\end{tabular}
}
\vspace{6pt}    
\caption{Catalogue of parameters used for the plots throughout this paper.}
\label{tab:plot_parameters}
\renewcommand{\arraystretch}{0.8}
\end{table}

\end{appendix}

\end{document}